\documentclass[aps,a4paper,preprint,eqsecnum,nofootinbib,groupedaddress,longbibliography,showpacs]{article}
\pdfoutput=1

\usepackage{jheppub}
\usepackage[utf8]{inputenc}

\usepackage{graphicx,epsf}
\usepackage{amsmath,amsfonts,amssymb}
\usepackage{yfonts}
\usepackage{xfrac}
\usepackage{bbold}
\usepackage{slashed}
\usepackage{graphicx}
\usepackage{multirow}
\usepackage{placeins}
\usepackage{wasysym}
\usepackage{physics}
\usepackage{mathtools}
\usepackage{tabularx}

\usepackage{enumitem}

\usepackage{natbib}
\usepackage{hyperref}
\usepackage{xcolor}
\hypersetup{
    colorlinks=true,
    allcolors={blue!60!black}
}

\usepackage[caption=false]{subfig}

\usepackage{etoolbox}
\preto\subequations{\ifhmode\unskip\fi}

\def\gmn{g^{\mu\nu}}
\def\g5{\gamma_5}

\newcommand\dt{D_{\text{t}}}
\newcommand\dtt{\widetilde{D}_{\text{t}}}
\newcommand\dttindex{$\dtt$-index}

\newcommand{\BH}{{\sc BlackHat}}

\providecommand\tr{}
\renewcommand\tr[2][]{
  \mathrm{Tr}_{#1} \left[ #2 \right]
}
\newcommand\trf[1][]{
  \mathrm{Tr} \left[ #1 \right]
}

\newcommand\trh[2][]{
  \widetilde{\mathrm{Tr}}_{#1} \left[ #2 \right]
}

\def\QS#1{\mathrm{QS}_{[#1]}}
\def\Sf{\mathrm{S}_{[4]}}

\def\gm#1#2#3{\prescript{(#3)}{}{\Gamma^{#1}_{[#2]}}}
\def\gml#1#2#3{\prescript{(#3)}{}{\Gamma^{\phantom{\mu}}_{#1\,[#2]}}}
\def\gmt#1#2{\prescript{(#2)}{}{\Gamma^{#1}}}
\def\Gm#1{\Gamma^{#1}}
\def\Gmds#1#2{\Gamma^{#1}_{[#2]}}

\def\pres#1#2{\prescript{(#2)}{}{#1}}

\newcommand\gtw[2]{\prescript{#2}{}{\textfrak{g}^{#1}}}
\newcommand\htw[2]{\prescript{#2}{}{\textfrak{h}^{#1}}}

\def\eqns#1#2{eqs.~(\ref{#1}) and~(\ref{#2})}

\def\Cl{\mathcal{C}\ell}

\begin{document}

\title{
  On the Dimensional Regularization of QCD Helicity Amplitudes With Quarks
}

\author[a]{F. R.~Anger}
\emailAdd{felix.anger@physik.uni-freiburg.de}
\author[a]{and V.~Sotnikov}
\emailAdd{vasily.sotnikov@physik.uni-freiburg.de}
\affiliation[a]{Physikalisches Institut, Albert-Ludwigs-Universit\"at Freiburg, D--79104 Freiburg, Germany}

%\arxivnumber{1803.xxxx}

\abstract{
We study QCD helicity amplitudes with an
arbitrary number of (massive) quarks, keeping unobserved (loop)
particles in fixed integer $D_s$ dimensions. We find a suitable
embedding of external four-dimensional fermion states into higher
dimensional spaces. This allows to identify the $D_s$ dependence of amplitudes with
external quarks at one and two loops, permitting an analytic
continuation in $D_s$. Explicitly we focus on 't Hooft-Veltman and four-dimensional helicity
schemes for which we provide a compact prescription for the
computation of one- and two-loop amplitudes amenable for numerical implementation.
}

\maketitle

\section{Introduction}

Dimensional regularization is a well established framework, which is used extensively in perturbative computations in quantum field theories. It is a key ingredient for precise predictions for
collisions of high energy particles at modern colliders. Various dimensional regularization schemes exist, which differ in their placement of observed (or external) particles and unobserved (or loop) particles in spaces with a dimensionality other than 4. For a recent review see \cite{Gnendiger:2017pys}.
A particular regularization scheme is typically chosen to provide the most convenient and efficient description
of the problem at hand. While analytic approaches are more flexible
when treating spin states in dimensional regularization, numerical
approaches to the computation of loop amplitudes, such as numerical
unitarity \cite{Ossola:2006us,Ellis:2007br,Giele:2008ve,Berger:2008sj},
are more limited and two conditions have to be considered:
\begin{enumerate}[label=\textbf{\roman*}]
  \item\label{cond1} The spaces on which the scheme is defined have to allow for the existence of well-defined finite-dimensional representations for all external and loop particles;
  \item\label{cond2}  The dimensionality of these spaces should be as small as possible to allow for an efficient implementation.
\end{enumerate}
The first condition is in contradiction with the analytic continuation
in the dimension parameter of loop-particle spin spaces $D_s$ away
from integer values used in dimensional regularization. We discuss this feature in the context of the `t Hooft-Veltman (HV) scheme \cite{tHooft:1972tcz} and the
four-dimensional helicity (FDH) scheme \cite{Bern:1991aq,Bern:2002zk}, which
keep external particles in four dimensions.
These two schemes differ in the dimensionality $D_s$. In HV it is set
to $D_s=4-2\epsilon$ and in FDH to $D_s=4$. We treat the two schemes uniformly by keeping
$D_s$ unspecified. While the extraction of the analytic dependence on $D_s$ has been
solved in the literature for gauge fields, the treatment of external
fermionic fields is incomplete and we provide a complete solution.

There exist two known approaches in numerical computations to get around the problem raised by condition \ref{cond1}.
Either some parts of the computation have to be performed with additional analytic input as in refs.~\cite{Ossola:2008xq,Fazio:2014xea,Bern:2010qa,Badger:2017jhb,Berger:2008sj},
or the (polynomial) dependence on $D_s$ has to be reconstructed through sampling over integer $D_s$, which we refer to as
dimensional
reconstruction \cite{Giele:2008ve,Ellis:2008ir,Boughezal:2011br}. In both cases the
scheme has to be complemented by a precise meaning of external particles being four-dimensional. This
is straightforward in pure Yang-Mills theory by embedding
four-dimensional polarization states into e.g.~the upper components of $D_s$ dimensional ones, but requires
some care when external fermions are present since the
embedding of four-dimensional states into $D_s$-dimensional Dirac
spinors is ambiguous. We propose a systematic prescription for the treatment of
external fermions in the HV/FDH dimensional regularization schemes based on the ideas of~\cite{Veltman:1988au}, which extends the
developments
of~\cite{Ellis:2008ir,DeFreitas:2004kmi,Bern:2010qa,Fazio:2014xea,Badger:2017gta}
in what regards the embedding of external fermion
states. We average over embeddings corresponding to a partial trace over the
higher-dimensional components of spinor chains. Our prescription is essential for any numerical multi-loop
computation of dimensionally regularized helicity amplitudes with external
fermions. Using only a single embedding of external fermions leads to
spurious integrand terms such as square roots of loop momenta
contractions. At one loop such terms are encountered  for
massive external fermions (see e.g. \cite{Fazio:2014xea,Badger:2017gta}), and beyond one loop also for massless
external fermions.
In proper analytic approaches to computations in the HV/FDH schemes, such as \cite{Cullen:2010jv}, these spurious integrand terms are not present.

In this work, we present a method to address the
conditions \ref{cond1} and \ref{cond2} simultaneously, thereby
allowing for an efficient numerical computation of
dimensionally regularized one- and two-loop amplitudes with
external quarks. We use explicit representations of the Clifford algebra
in integer $D_s$ dimensions, our embedding prescription for external fermions, and 
the tensor product structure of Clifford algebras in $D_s$ dimensions
to analytically identify the full $D_s$ dependence of one- and
two-loop QCD helicity amplitudes with an arbitrary number of external quarks.
The coefficients of polynomials in $D_s$ are expressed in terms of diagrams with modified particle content.
We then extrapolate $D_s$ to a desired value to obtain an amplitude in HV or FDH schemes.
We thereby extend the known decompositions by particle content for pure Yang-Mills theories at one and two
loops \cite{Bern:1994cg,Cheung:2009dc,Badger:2013gxa} to the full QCD
spectrum including massive external quarks. A direct application of dimensional reconstruction
techniques \cite{Giele:2008ve,Ellis:2008ir,Boughezal:2011br} for $n$-loop amplitudes requires computations in $n+1$ different (even) integer $D_s>4$ dimensions. 
Contrary to that, our approach uses only a single minimal dimension,
which is determined by the number of non-trivial directions of loop momenta.
This allows to alleviate the exponential growth%
\footnote{Representations of the Clifford algebra generated from $D_s$-dimensional space have the dimensionality of $2^{D_s/2}$.}
of the dimensionality of corresponding representations
for amplitudes with external quarks.

We validated our approach by implementing it for
the computation of dimensionally regulated one-loop amplitudes in a new version of the \BH{} library,
which we used to provide phenomenologically relevant
predictions for processes involving massive external
quarks \cite{Anger:2017glm}. In this work we consider bare color-ordered QCD
amplitudes. However since the argument presented relies mostly on the Lorentz
index structure of the particles involved, an extension to a broader class
of theories is possible.

This paper is organized as follows. In sec.~\ref{sec:FDH-def} we
review the considered regularization schemes. In sec.~\ref{sec:Ds-spin}
we establish the notation used in this work and state the properties of Clifford algebras and spinor states in $D_s$ dimensions.
In sec.~\ref{sec:helampl} we formulate the embedding prescription for external fermions.
We then work out the $D_s$ dependence of helicity amplitudes at one and two loops and
use it to obtain amplitudes in HV and FDH schemes.
In sec.~\ref{sec:fdf} we discuss the connection of our approach to the four-dimensional formulation (FDF) of
FDH \cite{Fazio:2014xea} at one loop. In sec.~\ref{sec:concl} we give
our conclusions. We provide implementation details of the computational prescription at
one loop in Appendix~\ref{sec:impl-deta}.

%%%%%%%%%%%%%%%%%%%%%%%%%%%%%%%%%%%%%%%%%%%%%%%%%%%
%%%%%%%%%%%%%%%%%%%%%%%%%%%%%%%%%%%%%%%%%%%%%%%%%%%

\section{FDH and HV Schemes}
\label{sec:FDH-def}

We follow the notation for different regularization schemes introduced in \cite{Gnendiger:2017pys}. It is based on distinguishing a four-dimensional Minkowski space $\Sf$, an infinite-dimensional
space of loop momentum integration $\QS{D}$ with $D=4-2\epsilon$ and an enlarged space ($D_s \geq D$) of internal spin states $\QS{D_s}$, such that
\begin{equation}
  \QS{D_s} = \QS{D} \oplus \QS{n_\epsilon} = \Sf \oplus \QS{n_\epsilon-2\epsilon}, \qquad n_\epsilon-2\epsilon = D_s-4.
  \label{eq:qssplit}
\end{equation}
We restrict our discussion to amplitudes computed in the FDH and HV schemes, that is all external particles are kept in $\Sf$, while loop particles are
extended to $\QS{D_s}$. These two schemes differ in the value of $D_s$, which we choose to keep general postponing
the specialization.  A split of tensor and Clifford algebras is induced by
\begin{subequations}
  \begin{align}
    g_{[D_s]}^{\mu\nu} &= g^{\mu\nu}_{[4]}+g^{\mu\nu}_{[D_s-4]},
    \end{align}
leading to the properties
      \begin{align}
     (g_{[\text{dim}]})^{\mu}_\mu
    &=\text{dim}, & (g_{[4]})^{\mu\rho}(g_{[D_s-4]})_{\rho\nu} &= 0,
  \end{align}
with dim $\in \{D_s,4,D_s-4\}$ and 
  \begin{align}
A^\mu_{[4]} &\equiv g^{\mu\nu}_{[4]} A^{\phantom{\mu}}_{\nu\,[D_s]},&  A^\mu_{[D_s-4]} &\equiv g^{\mu\nu}_{[D_s-4]} A^{\phantom{\mu}}_{\nu\,[D_s]},
  \end{align}
\end{subequations}
 where $A^{\mu}_{[D_s]}$ is any object carrying a vector index, like
 momenta or gamma matrices. We use
a left superscript in brackets to indicate the dimensionality $\dt$ of
the object explicitly, as in $\pres{A}{\dt}$, whenever there is a potential ambiguity.

The space $\QS{D_s}$ is formally infinite-dimensional for non-integer $D_s$. However, $D_s$ \textit{can} be chosen to be integer and
then $\text{dim}\left(\QS{D_s}\right) = D_s$. In this case an explicit
representation of tensor and spinor algebras is possible
(see e.g.~\cite{collins_1984}). After all sources of $D_s$ dependence are identified, an FDH or an HV
amplitude is obtained by setting $D_s=4$ or $D_s=4-2\epsilon$ respectively.

If four-dimensional external momenta are embedded trivially in $\QS{D_S}$, an integrand of a loop amplitude can only depend on
$(D-4)$ directions of loop momenta through contractions between themselves. Rotational symmetry of the loop
integration thus allows, without loss of generality, to write the loop momentum as
\begin{subequations}
  \label{eq:loopmom}
  \begin{align}
    \ell^\alpha_{[D]} &=   \ell^\alpha_{[4]} +   \ell^\alpha_{[D-4]} \equiv \ell^\alpha_{[4]} + i\sum_{j}\mu_j\,n_j^\alpha, \qquad n_j^2=1, \\
    \ell^2_{[D]} &= \ell^2_{[4]} + \ell^{2}_{[D-4]} \equiv \ell^2_{[4]}-\sum_j \mu_j^2,
  \end{align}
\end{subequations}
where $\{n^\alpha_j\}$ span non-trivial directions in $\QS{D-4}$, with $\text{dim}(\{n^\alpha_j\})$ given by the number of loops.

\section{Clifford Algebra and Spinors in Integer \texorpdfstring{$D_s$}{Ds}}
\label{sec:Ds-spin}
In this section, we review the definition of the objects of different dimensionality
that we handle in this work and the notation that we adopt for them. In
particular, we require representations of the
Clifford algebra and associated spinor states in arbitrary integer dimensions. 

\subsection{Representations of the Clifford Algebra}
\label{sec:repr-cliff-algebra}
A Clifford algebra $\Cl_{1,D_s-1}(\mathbb{R})$ generated from an even integer $D_s$-dimensional Minkowski space-time
with a signature $\{+,-,\ldots,-\}$ has a defining property
\begin{align}\label{eq:cliffordd}
  \{\Gamma^\mu,\Gamma^\nu\}=2\gmn,
\end{align}
where $\Gamma^\mu$ are the generating elements (which we refer to as
gamma matrices) and
$g^{\mu\nu}$ are the metric tensor components. Its $\dt=2^{Ds/2}$-dimensional representation can be constructed recursively \cite{collins_1984}:
\begin{subequations}
  \label{eq:gammaitall}
  \begin{align}\label{eq:gammait}
    \gmt{\mu}{\dt} \quad =& \pmqty{1 & 0 \\ 0 & 1} \otimes \gmt{\mu}{\frac{\dt}{2}}, & \mu &=0,\ldots,D_s-3,\\
    \label{eq:gammait2}
    & \pmqty{0 & i \\ i & 0} \otimes\gmt{\ast}{\frac{\dt}{2}}, & \mu &= D_s-2, \\
    \label{eq:gammait3}
    & \pmqty{0 & 1 \\ -1 & 0} \otimes \gmt{\ast}{\frac{\dt}{2}}, & \mu &= D_s-1
  \end{align}
with
\begin{align}
  \gmt{\ast}{\dt}=i^{\sfrac{D_s}{2}-1}\Gm{0}\cdots\Gm{D_s-1}, \qquad \Gm{\ast}\Gm{\ast}= \mathbb{1},
\end{align}
which can also be represented as
\begin{equation}
  \gmt{\ast}{\dt}= \pmqty{1 & 0 \\ 0 & -1}\otimes \gmt{\ast}{\frac{\dt}{2}}.
\end{equation}
\end{subequations}
We choose the four-dimensional Dirac $\gamma$-matrices as the starting point of the iteration:
\begin{align}
  \gmt{\mu}{4}& \equiv \gamma^\mu, &  \gmt{\ast}{4}& \equiv \g5.
\end{align}

It is useful to expose the tensor product structure of the Clifford algebra $\Cl_{1,D_s-1}\cong \Cl_{0,D_s-4}\otimes\Cl_{1,3}$.
The first four matrices in $D_s\geq 4$ obtained with the iteration in eq.~\eqref{eq:gammait}, can be written as
\begin{align}\label{eq:fourgamma}
  g^{\mu\nu}_{[4]} \gml{\nu}{D_s}{\dt} \equiv   \gm{\mu}{4}{\dt} =
  \begin{cases}
    \prescript{(\frac{\dt}{4})}{}{\mathbb{1}}\otimes \gamma^\mu, \qquad & \mu \leq 3, \\
    0, & \mu \geq 4,
  \end{cases}
\end{align}
that is they are $\dt\times \dt$ dimensional matrices with $\dt/4$ copies
of the four dimensional gamma matrices on the diagonal.
The remaining matrices can be written as
\begin{align}\label{eq:highgamma}
  g^{\mu\nu}_{[D_s-4]} \gml{\nu}{D_s}{\dt}   \equiv   \gm{\mu}{D_s-4}{\dt}  =  
  \begin{cases}
    0, & \mu \leq 3, \\
    \gtw{\mu}{(\frac{\dt}{4})}\otimes  {\g5}, \qquad & \mu \geq 4,
  \end{cases}
\end{align}
where $\gtw{\mu}{}$ are $(\sfrac{\dt}{4})$-dimensional matrices obtained from iteration of \eqns{eq:gammait2}{eq:gammait3}.
$\gtw{\mu}{}$ are generating elements of an Euclidean Clifford algebra $\Cl_{0,D_s-4}$.
We denote the dimensionality of its representation as $\dtt \equiv \dt/4$.

\subsection{Spinor States}
\label{sec:fermionic-states}
In $D_s$ dimensions the $\dt/2$ independent fermionic states $u_k$ must satisfy the polarization sum
\begin{align}\label{eq:qpolsumd}
  \sum_{k=1}^{\dt/2}u_{k}\bar{u}_{k}=  \slashed{\ell}+m,
\end{align}
as well as the Dirac equation
\begin{equation}\label{eq:diraceq}
  (\slashed{\ell}-m) u_k = 0, \qquad \bar{u}_{k}(\slashed{\ell}-m) = 0.
\end{equation}
The states constructed by the action of
the inverse Dirac operator on arbitrary reference spinors $q_k$ in $\dt$ dimensions
\begin{align}\label{eq:spconst}
  u_k(\ell,q_k) &= \frac{(\slashed{\ell}+m)}{\sqrt{N}}q_k, &  \bar{u}_k(\ell,q_k) &=\bar{q}_k\frac{(\slashed{\ell}+m)}{\sqrt{N}},
\end{align}
with a suitable normalization $N$ satisfy the requirements of \eqns{eq:qpolsumd}{eq:diraceq} by construction.
A \mbox{$\dt$-dimensional} basis for the reference spinors $q_k$ can be constructed recursively:
\begin{subequations}
  \label{eq:spinorsit}
  \begin{align}
    \prescript{(\dt)}{}{q_{k}} &= \pmqty{1\\0} \otimes   \prescript{(\frac{\dt}{2})}{}{q_{k}}, &   \prescript{(\dt)}{}{q_{k+\dt/4}} &=  \pmqty{0\\1}\otimes   \prescript{(\frac{\dt}{2})}{}{q_{k}}, \\
    \prescript{(\dt)}{}{\bar{q}_{k}} &= \pmqty{1 & 0}\otimes   \prescript{(\frac{\dt}{2})}{}{\bar{q}_{k}}, &   \prescript{(\dt)}{}{\bar{q}_{k+\dt/4}} &= \pmqty{0 & 1} \otimes   \prescript{(\frac{\dt}{2})}{}{\bar{q}_{k}},
  \end{align}
\end{subequations}
with $k\in\{0,\ldots ,\dt/4-1\}$. We choose as the
starting point of this recursive definition in $D_s=4$ the Weyl spinors
\begin{subequations}
  \begin{align}
    \prescript{(4)}{}{q_{1}}&\equiv\vert q,+\rangle,  &
  \prescript{(4)}{}{q_{2}}&\equiv\vert q,-\rangle,\\
  \prescript{(4)}{}{\bar{q}_{1}}&\equiv\langle q,+\vert  , &
    \prescript{(4)}{}{\bar{q}_{2}}&\equiv\langle q,-\vert , 
  \end{align}
\end{subequations}
representing states with definite helicity. 

The $\dt$-dimensional spinors for four-dimensional momenta have a
particularly simple form. To reflect the diagonal structure of the $\dtt$-dimensional
part of the first four gamma matrices in eq.~\eqref{eq:fourgamma}
we split the index $k$ labeling $\dt/2$ states into an index $i\in\{0,1\}$ denoting the helicities $\{+,-\}$ and an index
$j\in\{0,\ldots, \dtt-1\}$ such that $k=i+2j$ and obtain
\begin{align}\label{eq:4dst}
    \prescript{(\dt)}{}{u_{i,j}(p_{[4]})} &= \vert j\rangle \otimes \vert p,\pm\rangle \equiv \vert p,\pm,j\rangle,\notag \\
    \prescript{(\dt)}{}{\bar{u}_{i,j}(p_{[4]})} &= \langle j\vert  \otimes \langle p,\pm\vert  \equiv \langle p,\pm,j\vert , &j \in \{0,\ldots,\dtt-1\}, 
\end{align}
with $\{\vert j \rangle \} $ forming an orthonormal basis in the $\dtt$-dimensional space corresponding to the representation of
an Euclidian Clifford algebra, such that
\begin{align}
  \langle p_1,\pm,j_1 \vert  p_2,\pm,j_2 \rangle =   \langle p_1,\pm \vert 
  p_2,\pm \rangle \delta_{j_1j_2}.
\end{align}
We refer to the index $j$ of the fermion states as the
\dttindex{}. Note that there are in general $\dtt$ states for each helicity choice in $D_s$ dimensions.

\subsection{Normalized Partial Traces}
\label{subsec:trtw}

In this work we make use of normalized partial traces over the subspace $\dtt$ of products of matrices
$\Gmds{\mu}{D_s}$, which we define as a map
\begin{equation}
  \trh{\cdot} : (\Cl_{0,D_s-4}\otimes\Cl_{1,3}) \mapsto \Cl_{1,3},
  \label{eq:trdef}
\end{equation}
such that for all $\tilde{A}\in \Cl_{0,D_s-4}$ and  $B\in \Cl_{1,3}$
\begin{align}
  \trh{\tilde{A} \otimes B} = \frac{1}{\dtt} \Tr[\tilde{A}]\cdot B,
  \label{eq:trdefprop}
\end{align}
where $\Tr[\tilde{A}]$ is a usual trace in $\Cl_{0,D_s-4}$. Given the recursive construction by eqs.~\eqref{eq:gammaitall} and \eqref{eq:spinorsit},
it can be explicitely evaluated as
\begin{equation}
  \label{eq:explicittrace}
  \pres{A}{4}_{ij} \equiv \left(\trh{\pres{A}{\dt}}\right)_{ij} = \frac{1}{\dtt}\,\sum_{k=1}^{D_s-4} \pres{A}{\dt}_{i+4(k-1),\,j+4(k-1)},
\end{equation}
where $\pres{A}{4}_{ij}$ are the components of the image of $\pres{A}{\dt}_{ij}$. This can be interpreted
as an averaging over all possible embeddings of the four-dimensional states into a $D_t$-dimensional representation.
We present some properties of normalized partial traces in the Appendix \ref{sec:trace-properties}.

The definition given by eq.~\eqref{eq:trdef} can be generalized for any choice of a tensor product split $\Cl_{0,D_s-x}\otimes\Cl_{1,x-1}$
with $4\leq x\leq D_s-2$. In particular we use the case with $x=6$ for two-loop computations in section \ref{sec:dsdep-2loop}.

\section{Helicity Amplitudes}
\label{sec:helampl}

We propose to compute integrands of dimensionally regularized loop helicity amplitudes in HV or FDH schemes using finite-dimensional
representations of the Clifford algebra. The building blocks of our
approach are amplitudes with loop particles in $\QS{D_s}$ and four-dimensional momenta and polarization states of external particles
embedded in the enlarged $D_s$-dimensional space. 

In the first part of this section, we work out the details of a correct treatment
of external particles in the considered regularization schemes. 
In the second part, we then analyze the $D_s$ dependence of amplitudes
at one and two loops. The resulting decomposition by particle
content allows to compute in spaces of reduced dimensionality and HV or FDH amplitudes are obtained by extrapolating $D_s$ to the corresponding value.

\subsection{Embedding Prescription}
\label{sec:olmetrace}

Given a representation of a Clifford algebra $\Cl_{1,D_s-1}(\mathbb{R})$ in $\dt = 2^{D_s/2}$ dimensions,
one can construct a basis of $2^{D_s}$ elements
\begin{align}
  \label{eq:clbasis}
  &\left\{
    \mathbb{1}^{\phantom{[\mu]}},\Gm{[\nu_1]},\Gm{[\nu_1\nu_2]},\ldots,
    \Gm{[\nu_1\ldots\nu_{D_s}]}\right\},& \nu_i \in \{0,\ldots,D_s-1\},
\end{align}
from antisymmetric combinations of $\Gamma^\nu$ as follows:
\begin{align}
  \Gm{[\nu_1\ldots\nu_m]}\equiv \frac{1}{m!}\sum_{\sigma \in
    \mathcal{P}_m} \text{sgn}(\sigma) \Gm{\sigma_1}\cdots \Gm{\sigma_m},
\end{align}
where the summation runs over all permutations and the parity
$\text{sgn}(\sigma)$ of each
permutation is given by the number of inversions in it and evaluates to $+1$ or $-1$ for
even or odd permutations respectively.

Any product of gamma matrices
\begin{equation}\label{eq:gammastring}
  S^{\mu_1\ldots \mu_n} = \Gm{\mu_1}\cdots \Gm{\mu_n}
\end{equation}
can be decomposed using the basis \eqref{eq:clbasis} as
\begin{align}
  S^{\mu_1\ldots \mu_n} = \alpha^{\mu_1\ldots \mu_n} \mathbb{1} +
  \alpha_{\nu_1}^{\mu_1\ldots \mu_n} \Gm{[\nu_1]} +
  \alpha_{\nu_1\nu_2}^{\mu_1\ldots \mu_n} \Gm{[\nu_1\nu_2]}+\ldots+\alpha_{\nu_1\cdots\nu_{D_s}}^{\mu_1\ldots \mu_n} \Gm{[\nu_1\cdots\nu_{D_s}]}.
\end{align}
For the special case of $D_s=4$ we have:\footnote{%
The basis in four dimensions is typically simplified making use of $\gamma_5$, however
this is not advantageous for our argument.
}
\begin{align}
  S^{\mu_1\ldots \mu_n}_{[4]} = \alpha^{\mu_1\ldots \mu_n} \mathbb{1} +
  \alpha_{\nu_1}^{\mu_1\ldots \mu_n} \Gm{[\nu_1]} +
  \alpha_{\nu_1\nu_2}^{\mu_1\ldots \mu_n} \Gm{[\nu_1\nu_2]}+\alpha_{\nu_1\nu_2\nu_3}^{\mu_1\ldots \mu_n} \Gm{[\nu_1\nu_2\nu_3]}+\alpha_{\nu_1\nu_2\nu_3\nu_4}^{\mu_1\ldots \mu_n} \Gm{[\nu_1\nu_2\nu_3\nu_4]}.
\end{align}
Using the property $\gm{\mu}{4}{\dt} =
\pres{\mathbb{1}}{\dtt}\otimes\gamma^{\mu}$ (see
eq.~\eqref{eq:fourgamma}) and the normalized partial trace defined in
section \ref{subsec:trtw} we can identify the basis elements of the
four-dimensional basis embedded trivially in $\dt$ dimensions by
\begin{align}\label{eq:embedldhd}
  \gmt{[\nu_1\cdots \nu_m]}{4} = \trh{\gm{[\nu_1\cdots\nu_m]}{4}{\dt}}.
\end{align}

An important observation is that the basis elements $\Gmds{[\nu_1\cdots\nu_m]}{4}$ form a subset of the
$\dt$-dimensional basis and we can thus split the basis decomposition of eq.~\eqref{eq:gammastring}
 in $\dt$ dimensions into a direct sum
\begin{align}\label{eq:expslplit}
    S^{\mu_1\ldots \mu_n} =
    S_{[4]}^{\mu_1\ldots \mu_n} \oplus S_{[D_s-4]}^{\mu_1\ldots \mu_n}.
\end{align}

 A well-defined treatment of loop helicity amplitudes in
integer dimensions in the FDH or HV scheme requires external
particle states to be in four-dimensions, cf.~discussion in Sec.~\ref{sec:FDH-def}. Conversely, gamma matrices originating in
interactions of loop particles are forced to be $D_t$-dimensional. Therefore, the spinors
of external particles need to be embedded in the $\dt$-dimensional
space. To this end, taking inspiration from the ideas of \cite{Veltman:1988au}, we would have to require that
the states $\pres{u_{\pm}}{\dt}$ and $\pres{\bar{u}_{\pm}}{\dt}$ representing external particles with definite helicity
satisfy
\begin{subequations}
    \label{eq:reqtrem}
  \begin{align}
    \label{eq:reqtrem1}
    \prescript{(\dt)}{}{ \bar{u}_{\pm}^{\phantom{[}}} \gm{[\mu_1\ldots\mu_n]}{D_s-4}{\dt}
      \prescript{(\dt)}{}{u_{\pm}^{\phantom{[}}} \overset{!}{=}~&0,&
        \forall\  \gm{[\mu_1\ldots\mu_n]}{D_s-4}{\dt}&\in \prescript{(\dt)}{}{S_{[D_s-4]}^{\mu_1\ldots \mu_n}}, \\
      \label{eq:reqtrem2}
      \prescript{(\dt)}{}{ \bar{u}_{\pm}^{\phantom{[}}} \gm{[\mu_1\ldots\mu_n]}{4}{\dt}
      \prescript{(\dt)}{}{u_{\pm}^{\phantom{[}}} \overset{!}{=}~&
        \prescript{(4)}{}{ \bar{u}_{\pm}^{\phantom{[}}} \gm{[\mu_1\ldots\mu_n]}{4}{4}
        \prescript{(4)}{}{u_{\pm}^{\phantom{[}}},&
        \forall\  \gm{[\mu_1\ldots\mu_n]}{4}{\dt}&\in \prescript{(\dt)}{}{S_{[4]}^{\mu_1\ldots \mu_n}},
    \end{align}
\end{subequations}
that is all matrix elements of four-dimensional states trivially embedded in a higher-dimensional space annihilate
all $[D_s-4]$ basis elements in \eqref{eq:expslplit} while leaving all four-dimensional basis elements unchanged.
However one can show that no such states can be found in finite-dimensional representations.
To overcome this we propose a consistent prescription in which the conditions of eq.~\eqref{eq:reqtrem} are fulfilled by a normalized partial trace over
$\dtt$,
\begin{multline}
  \prescript{(\dt)}{}{ \bar{u}_{\pm}}~\pres{S^{\mu_1\ldots \mu_n}}{\dt}
  \prescript{(\dt)}{}{u_{\pm}} \iff \\ 
  \sum_{j=1}^{\dtt} \left\langle p_1,\pm,j \right\vert~\pres{S^{\mu_1\ldots \mu_n}}{\dt} \left\vert p_2,\pm,j \right\rangle~\equiv~
  \left\langle p_1,\pm \right\vert~\trh{\pres{S^{\mu_1\ldots \mu_n}}{\dt}} \left\vert p_2,\pm \right\rangle.
  \label{eq:main-prescription}
\end{multline}
Indeed, we have
\begin{subequations}
  \begin{align}
    \trh{  \prescript{(\dt)}{}{S_{[D_s-4]}^{\mu_1\ldots \mu_n}}}&=0\label{eq:trreq}\\
    \trh{  \prescript{(\dt)}{}{S_{[4]}^{\mu_1\ldots
    \mu_n}}}&= \prescript{(4)}{}{S_{[4]}^{\mu_1\ldots
    \mu_n}},
    \label{eq:trreq2}
  \end{align}
\end{subequations}
which correspond to \eqns{eq:reqtrem1}{eq:reqtrem2} respectively.
Eq.~\eqref{eq:trreq} follows from the fact that basis elements are either
antisymmetric in the $\dtt$ space with respect to at least a single
pair of indices or contain a single gamma
matrix with $\mu \geq 4$. In both cases, the basis elements are eliminated by
the partial trace. The relation in
eq.~\eqref{eq:trreq2} follows from the defining property of the partial trace, see eq.~\eqref{eq:trdefprop}.

Thus the normalized partial trace over the $\dtt$ space allows a consistent
embedding of four-dimensional external fermion states into a $D_s$-dimensional space.
This can be combined with an embedding of four-dimensional vector-boson states into a $D_s$-dimensional space,
which does not pose any difficulty.
An \emph{integrand} of a loop helicity amplitude with all external particles embedded trivially is then schematically given by
\begin{align}
  \label{eq:trampl}
  \mathcal{A}^{[D_s]}_{\text{external}\,\in\,S_{[4]}}~=~ \widetilde{\mathrm{Tr}}_{l_1} \cdots \widetilde{\mathrm{Tr}}_{l_n}\left[ \mathcal{A}^{[D_s]}(j_{l_1},\ldots,j_{l_n}) \right],
\end{align}
where $\mathrm{Tr}_{l_i}$ denotes the substitution of eq.~\eqref{eq:main-prescription} applied to a pair of external fermions $i$.
Note that $\mathcal{A}^{[D_s]}(j_{l_1},\ldots,j_{l_n})$ is an amplitude in $D_s$ dimensions with external fermions having open $\dtt$-indices $j_{l_i}$,
which means that eq.~\eqref{eq:trampl} defines a map from $\dt/2$ degrees of freedom per fermion to two degrees of freedom (two helicities) per fermion.

Amplitudes with multiple identical external quark-antiquark pairs can be decomposed into a properly antisymmetrized sums of amplitudes with distinct quark-antiquark pairs.
Our embedding prescription is then applied to each of the amplitudes with distinct quark-antiquark pairs separately.

\subsection{\texorpdfstring{$D_s$}{Ds} Dependence at One Loop}
\label{sec:dsdep-1loop}

We now proceed to establish the explicit $D_s$ dependence of one-loop amplitudes by analyzing all possible non-trivial
contributions to traces over $\dtt$ in eq.~\eqref{eq:trampl}. They can be
grouped in four different classes depending on the type of particles in the loop: closed gluon loops, one fermion line entering the loop,
two or more fermion lines entering the loop, and closed fermion loops.

In the following, we use
the split of an axial gauge\footnote{A similar, but more involved
  argument can be carried out with the Feynman gauge
  propagator.} gluon propagator in $D_s$ dimensions
induced by a specific choice of the reference momentum (see Appendix
\ref{sec:glue-axial}). Since also gluonic vertices respect this
split-up, a $D_s$-dimensional gluon line in the loop decomposes into a four dimensional, ``massive'' gluon with a mass
$\mu^2$ and ($D_s-5$) scalar components. The decomposition of pure gluon amplitudes by particle content, with the
split-up into a
``massive'' gluon amplitude in four-dimensions\footnote{Or equivalently a massless
  gluon in five-dimensions.} and a scalar amplitude,
was previously established in \cite{Bern:1994cg}. In the remainder of this
section, we establish a decomposition valid for the full QCD spectrum including several (massive) quark lines.

\begin{figure}[h]%{L}{0.45\textwidth}
  \centering
  \includegraphics[width=0.30\textwidth]{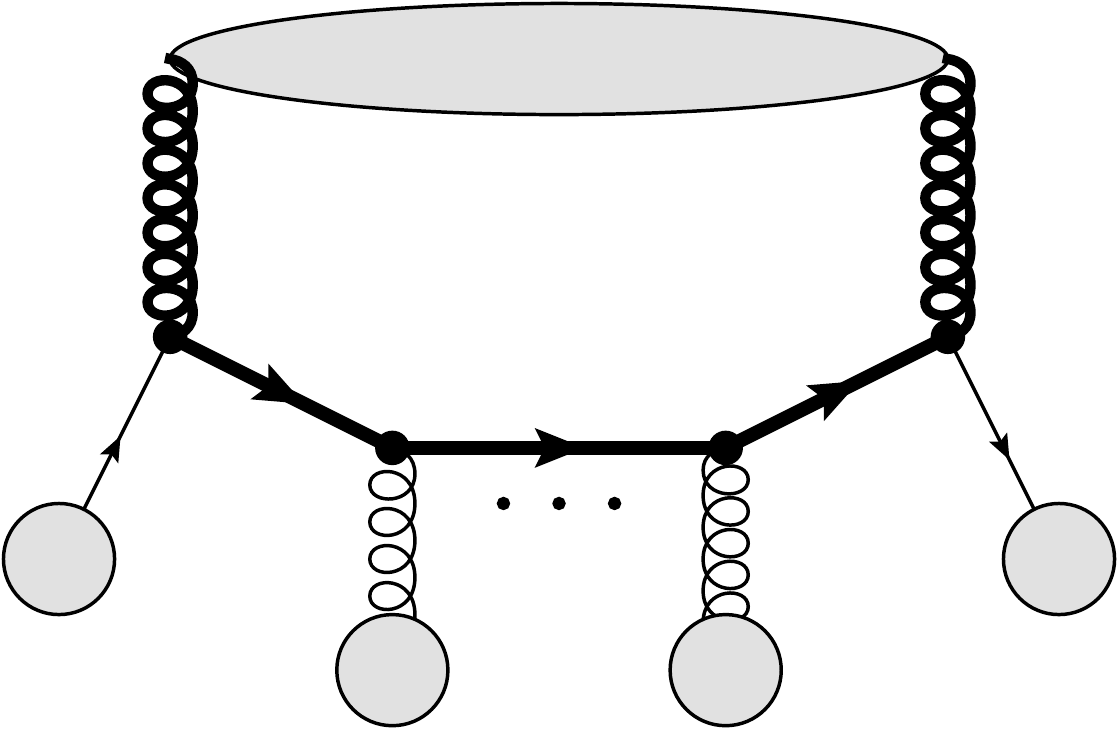}
  \caption{A schematic subdiagram for a contribution from a generic spinor line entering the loop. Bold lines and vertices represent
  objects in $D_s$ dimensions. Blobs represent parts of the diagram which explicit form is not important here.}
  \label{fig:spinchain}
\end{figure}

When fermion lines enter (and leave) the loop, they
contribute generic spinor chains to diagrams
(Fig.~\ref{fig:spinchain}). The source of gamma matrices contributing
to the trace over $\dtt$ is twofold. Vertices of quarks coupled to
gluons carrying loop momentum are proportional to $\Gmds{\mu}{D_s}$.
Propagators of quarks with mass $m$ carrying loop momentum $\ell$ can be written, using relations \eqref{eq:loopmom}, as
\begin{equation}\label{eq:qprop}
   S_{\text{q}}(\ell_{[D]}) = i~\frac{\ell^{\phantom{\mu}}_{[D]\,\alpha}\Gm{\alpha}+m\cdot\mathbb{1}}{\ell^2_{[D]}-m^2} =
  \frac{i}{\ell^2_{[4]}-\mu^2-m^2}\left( \slashed{\ell}_{[4]} +m\cdot\mathbb{1} + \mu\,\Gm{4} \right),
\end{equation}
where the only non-trivial in $\dtt$ contribution comes from $\mu\,\Gm{4}$,
since the first two terms inside the brackets are proportional to a
unit matrix in $\dtt$. A generic spinor string $T^{\mu\nu}_{[D_s]}$, with indices coupled to $D_s$-dimensional
gluons in the loop left uncontracted, is then of the form
\begin{equation}\label{eq:protochain}
  T^{\mu\nu}_{[D_s]} =
  \langle p_1, \pm \vert~\trh{ \cdots \Gm{\mu}\cdots
  \left( \prod_{k=0}^{n}\Gm{4} \cdots \right)\cdots  \Gm{\nu}\cdots }~| p_2, \pm \rangle,
\end{equation}
where the dots $\cdots$ represent any number of
gamma matrices which contribute only a unit matrix to the trace.
The partial trace can be evaluated explicitly using \eqns{eq:fourgamma}{eq:highgamma}.
It is convenient  to absorb the sign stemming
from the square of even powers of $\gtw{4}{}$ insertions from quark
propagators into the remaining four-dimensional $\g5$ by
\begin{align}\label{eq:tensdecomp2a}
  \ell^{\phantom\mu}_{\alpha\,[D_s-4]}  \Gmds{\alpha}{D_s-4}   &=
  \mu\left(\gtw{4}{}\otimes  {\g5}\right) \equiv  \mu\left(-i\gtw{4}{}\right)\otimes (i {\g5}),
\end{align}
such that the we have $(-i\gtw{4}{})^2=1$.
Using the fact that $T^{\mu\nu}_{[D_s]}$ is only non-vanishing for an even number of $\Gmds{\mu}{D_s-4}$ insertions
and trace properties from Appendix \ref{sec:trace-properties} we can evaluate it to
\begin{equation}\label{eq:traceprop}
  T^{\mu\nu}_{[D_s]}~=~%\trh{\cdots \Gmds{\mu}{D_s}\cdots\left(\prod_{k=0}^{n}\Gmds{4}{D_s}\cdots\right)\cdots\Gmds{\nu}{D_s}\cdots} = \\
    \begin{cases}
      \big(\cdots \gamma^{\mu}\cdots\gamma^{\nu}\cdots\big) ,                                 &\mu,\nu \leq 3,~n\text{ even}, \\
      \big(\cdots \g5\cdots\g5\cdots\big) ~ g_{[D_s-4]}^{\mu\nu},            &\mu,\nu \geq 4,~n\text{ even}, \\
      -i~\big(\cdots \g5\cdots\gamma^{\nu}\cdots\big) ~ g_{[D_s-4]}^{\mu 4 },      &\mu \geq 4, \nu \leq 3,~n\text{ odd}, \\
      -i~\big(\cdots \gamma^{\mu}\cdots\g5\cdots\big) ~ g_{[D_s-4]}^{\nu 4 }, \quad  &\mu \leq 3, \nu \geq 4,~n\text{ odd},
    \end{cases}
\end{equation}
where only the remainders of $\dt$-dimensional gamma matrices with
open indices from eq.~\eqref{eq:protochain} are explicitly shown, and the contraction
with external states is implicit.

\begin{figure}[h]
    \centering
    \subfloat[][\label{fig:1l_2q}one quark line]{ \includegraphics[height=16ex]{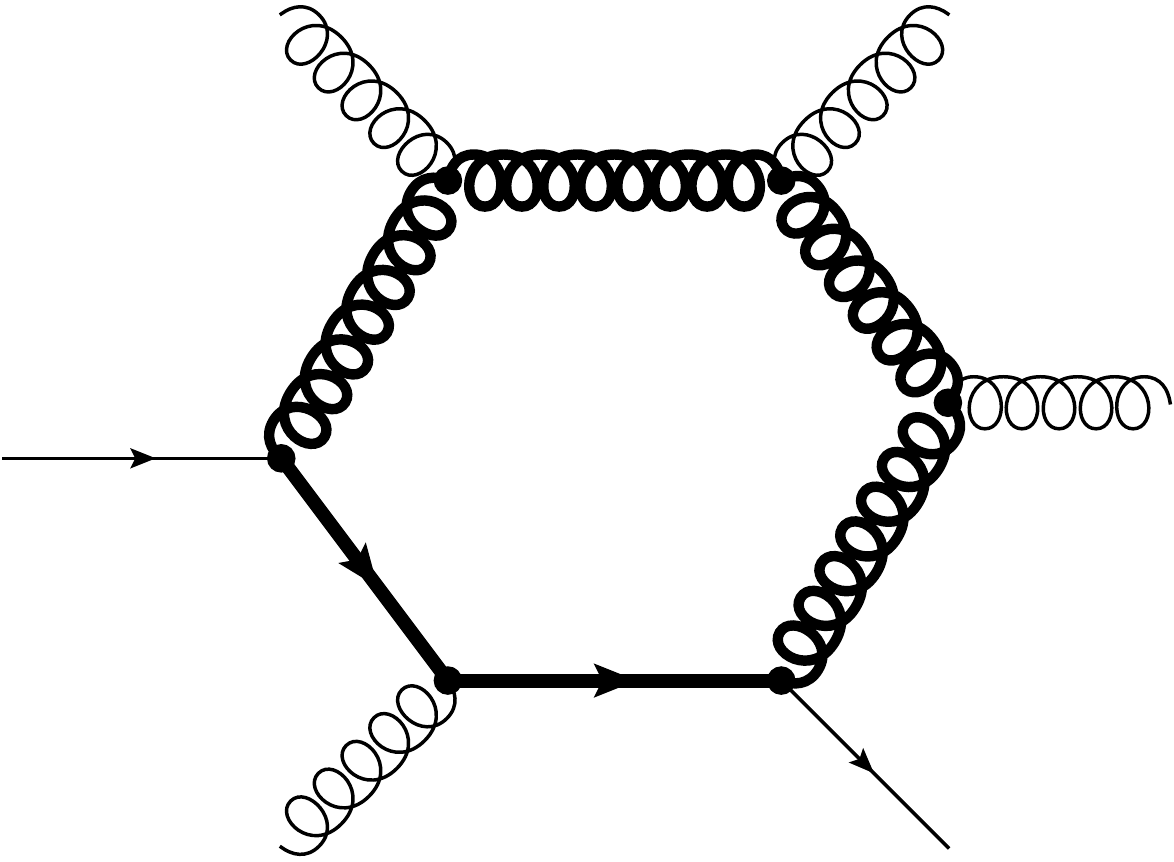}}
    \hspace{1cm}
    \subfloat[][\label{fig:1l_4q}two quark lines]{ \includegraphics[height=16ex]{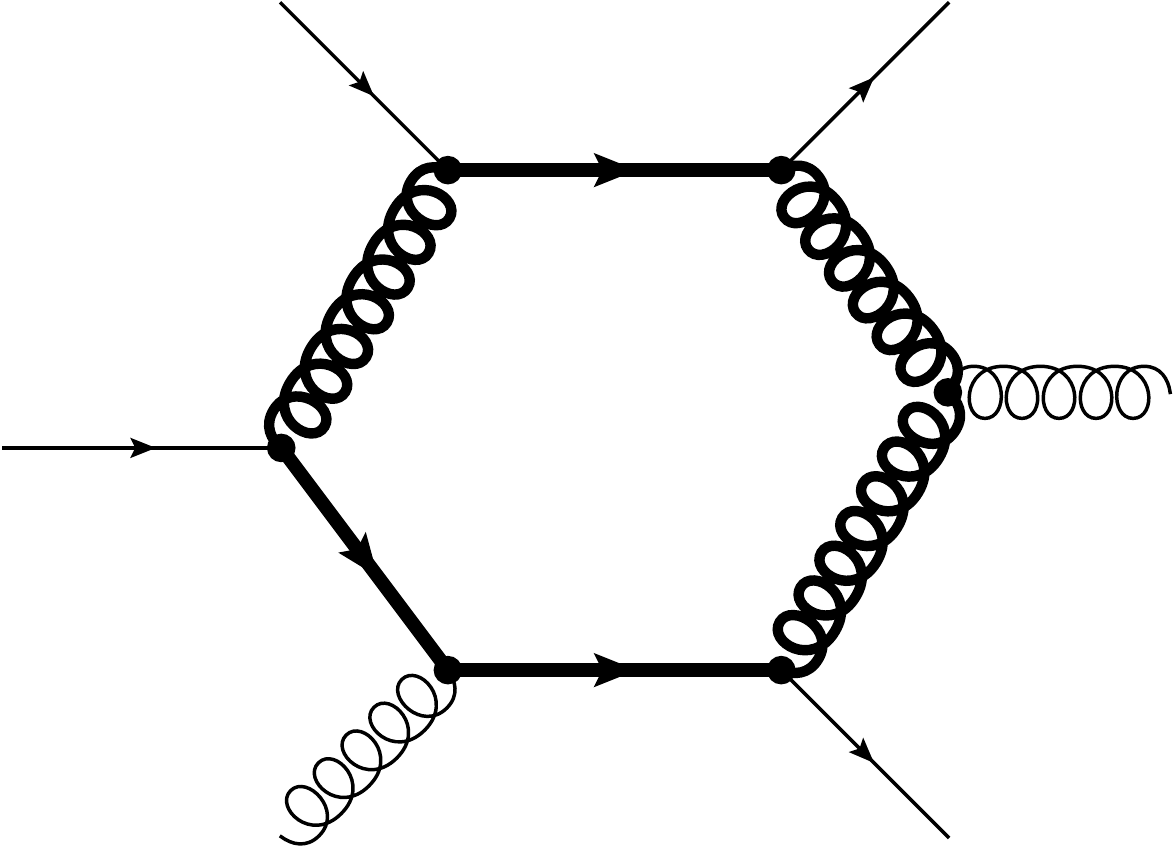}}
    \hspace{1cm}
    \subfloat[][\label{fig:1l_nf}closed quark loop]{ \includegraphics[height=16ex]{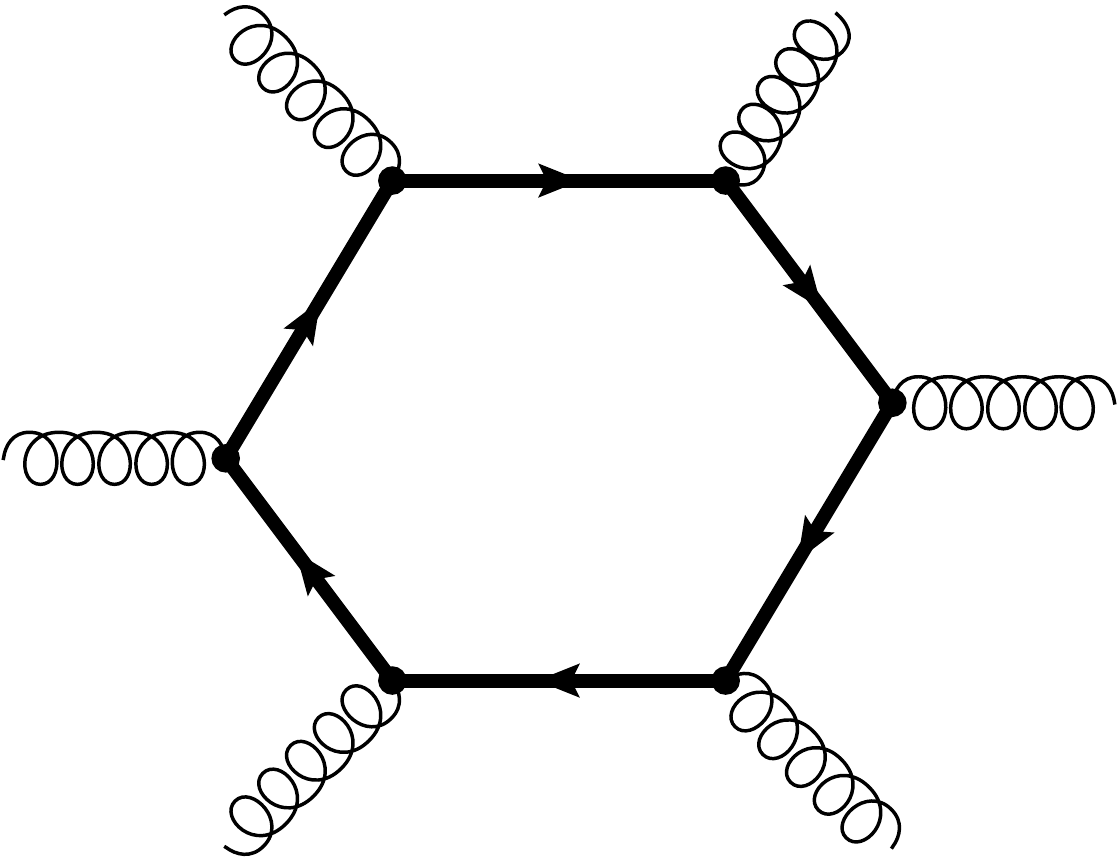}}
    \caption{Examples of diagrams with quarks in the loop. Bold lines and vertices represent objects in $D_s$ dimensions.}
    \label{fig:1l_examples}
\end{figure}

The result of eq.~\eqref{eq:traceprop} then needs to be contracted with the rest of the loop diagram $D^{\mu\nu}_{[D_s]}$.
If only a single fermion line enters the loop (e.g. as in fig.~\ref{fig:1l_2q}), we can use the decoupling of ``massive'' gluons and scalars and write
\begin{equation}
  D^{\phantom{\mu}}_{[D_s]\,\mu\nu} = \{\cdots\}_{[4]\,\mu\nu} + \{\cdots\}^{\phantom{\mu}}_\mathcal{S} \cdot g^{\phantom{\mu}}_{[D_s-5]\,\mu\nu},
\end{equation}
where $\{\cdots\}_{[4]\,\mu\nu}$ is a tensor with all $D_s$-dimensional gluons replaced by ``massive'' gluons
and $\{\cdots\}_\mathcal{S}$ is a contribution with all $D_s$-dimensional gluons replaced by scalars.
Thus the contraction takes the form
\begin{equation} \label{eq:contr1}
  T^{\mu\nu}_{[D_s]} D^{\phantom{\mu}}_{[D_s]\,\mu\nu} =
  T^{\mu\nu}_{[D_s]}\left( \{\cdots\}_{[4]\,\mu\nu} + \{\cdots\}^{\phantom{\mu}}_S \cdot g^{\phantom{\mu}}_{[D_s-5]\,\mu\nu} \right)=
  d_{\mathcal{G}} + (D_s-5)\,d_{\mathcal{S}},
\end{equation}
where $d_{\mathcal{G}}$ is a diagram with ``massive'' gluons
and $d_{\mathcal{S}}$ is a diagram with scalars. In the second equality we used \eqref{eq:traceprop} and the property
\begin{equation} \label{eq:metrconr}
  g^{\mu\nu}_{[D_s-4]}g_{[D_s-5]\,\nu\mu}^{\phantom{\mu}}=g^\mu_{\mu\,[D_s-5]}=(D_s-5).
\end{equation}

For loop diagrams with multiple quark lines entering the loop (e.g. as in fig.~\ref{fig:1l_4q}), the
split-up of $D_s$-dimensional gluons leads to mixed diagrams with both
``massive'' gluons and scalars coupling to the same spinor string which have to be considered in general.
However the contraction of a fermion line with a ``massive'' gluon on one side and
with a scalar on the other side turns out to vanish. Indeed, in this case the last two
cases from eq.~\eqref{eq:traceprop} are chosen, which leads to
\begin{equation}
  g_{[D_s-5]}^{\mu \nu  } g_{[D_s-4]\, \nu 4}^{\phantom{\mu}} = g_{[D_s-5]}^{\mu 4} = 0.
\end{equation}
Therefore, diagrams with ``massive'' gluons in the loop and those with
scalars in the loop decouple in the case of multiple quark lines as well. The
diagrams with scalars in the loop pick up a factor of $(D_s-5)$
from contractions across multiple lines as in eq.~\eqref{eq:metrconr}.

Summing up, we found that all diagrams exhibit the same dimensional dependence and respect the decoupling of ``massive'' gluons and scalars.
Hence a partially traced amplitude with external particles in $S_{[4]}$ and any number of external
quark pairs can be written in arbitrary $D_s$ as
\begin{equation}
  \mathcal{A}^{[D_s]}_{\text{external}\,\in\,S_{[4]}} =  \mathcal{A}_{\mathcal{G}} + (D_s-5)~\mathcal{A}_{\mathcal{S}},
  \label{eq:ddependence}
\end{equation}
where $\mathcal{A}_{\mathcal{G}}$ is a sum of diagrams with ``massive''
gluons in the loop and $\mathcal{A}_{\mathcal{S}}$ is a sum of diagrams with
scalars in the loop. Both $\mathcal{A}_{\mathcal{G}}$ and $\mathcal{A}_{\mathcal{S}}$ are defined in terms of four-dimensional objects with the additional
requirement of only even number of $(D_s-4)$-dimensional loop momentum insertions from the quark propagator \eqref{eq:qprop} on \emph{each}
fermion line. The latter is in general impossible to ensure numerically
\emph{only} with four-dimensional representations of the Clifford algebra, hence it is the ultimate obstruction
to such an attempt. 
Further discussion and details for the computation of $\mathcal{A}_{\mathcal{G}}$ and  $\mathcal{A}_{\mathcal{S}}$ are provided in Appendix~\ref{sec:impl-deta}.

Once the full dimensional dependence is established by eq.~\eqref{eq:ddependence}, we can use it to \emph{define}
an amplitude for continuous $D_s$.
An FDH or HV amplitude can be then obtained by setting $D_s=4$ or to $D_s=4-2\epsilon$ correspondingly:
\begin{subequations}
  \begin{align} \label{eq:decgsc}
    \mathcal{A}^{\text{FDH}} &= \mathcal{A}_{\mathcal{G}} - \mathcal{A}_{\mathcal{S}}, \\
    \mathcal{A}^{\text{HV}} &= \mathcal{A}_{\mathcal{G}} - (1+2\epsilon) \mathcal{A}_{\mathcal{S}} = \mathcal{A}^{\text{FDH}} - 2\epsilon \mathcal{A}_{\mathcal{S}}.
  \end{align}
\end{subequations}

Finally, diagrams with a closed fermion loop (e.g. as in fig.~\ref{fig:1l_nf}) do not have any $D_s$-dimensional vector index contractions, hence the only
possible dimensional dependence is an additional total factor of $\dtt$ from the loop trace, which we can cancel explicitly and get
\begin{equation} \label{eq:nfdecomp}
  \mathcal{A}^{\text{FDH}}_{\text{nf}} = \mathcal{A}^{\text{HV}}_{\text{nf}} = \frac{1}{\dtt}~\mathcal{A}^{[D_s]}_{\text{nf}}.
\end{equation}

We validated our approach by comparing NLO virtual matrix elements of most QCD
processes with up to 7 partons including massive top and bottom quarks as well
as those with an associated emission of a $W^{\pm}$ boson%
\footnote{
  Chiral couplings can be incorporated using the t'Hooft-Veltman prescription, see Appendix~\ref{subsec:g5}.
}
with publicly available tools.

\newlength{\Dht}
\setlength{\Dht}{13ex} %height of diagrams

\subsection{\texorpdfstring{$D_s$}{Ds} Dependence at Two Loops}
\label{sec:dsdep-2loop}
The results of sections \ref{sec:FDH-def}, \ref{sec:Ds-spin}, and \ref{sec:olmetrace}, in particular eq.~\eqref{eq:trampl}, are valid at any loop order.
However the analytical evaluation of partial traces that we accomplished in the previous section is not possible beyond one loop.
Considering a two-loop computation we perform a two-step process: partial traces over
dimensions higher than six can be performed analytically once and for all, and the
remaining traces have to be computed explicitly.

\begin{figure}[h]
    \centering
    \subfloat[][\label{fig:D2generic}generic topology with quarks in loops]{ \includegraphics[height=\Dht]{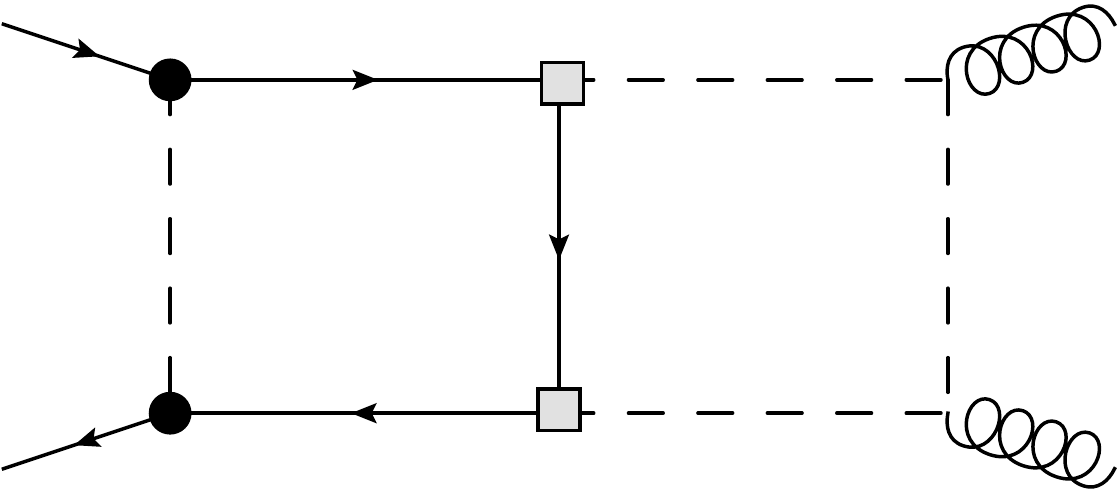}}
    \hspace{2cm}
    \subfloat[][\label{fig:facttopology}factorized topology]{ \includegraphics[height=\Dht]{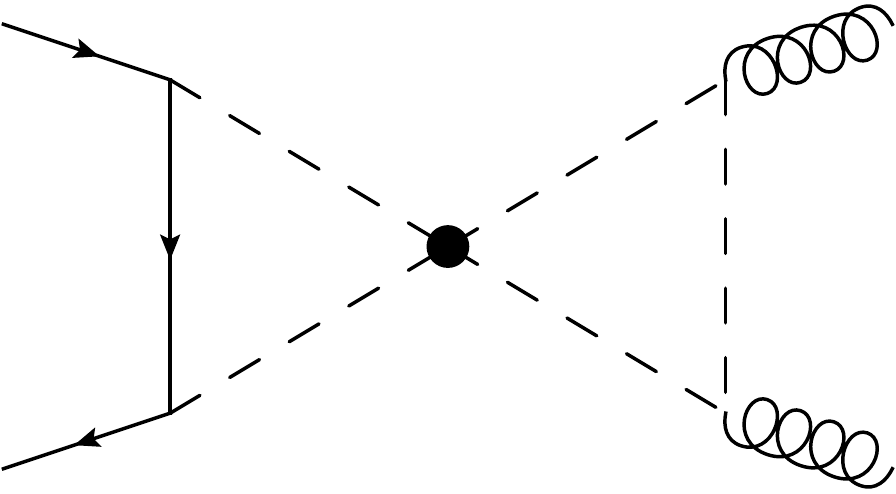}}
    \caption{Diagrams contributing to the $(D_s-6)^2$ coefficient. Quark-scalar vertices originating from equal indices are represented by the same shape.}
\label{fig:D2}
\end{figure}

At two loops we need at least $D_s=6$
to embed two non-trivial directions of loop momenta (see eq.~\eqref{eq:loopmom}).
We split $D_s$-dimensional gluons in six-dimensional (6d) gluons and scalars.
Similar to the one-loop case, scalar lines start and end only on quarks lines, or they form a closed loop otherwise (see Appendix \ref{sec:glue-axial}).
At one loop we chose an axial
gauge to effectively reduce five-dimensional gluon to a four-dimensional ``massive'' gluon and keep quark propagators in four dimensions.
Since the dimensionality of the loop momentum is even at two loops,
such a decomposition is not beneficial. For pure Yang-Mills theory the dimensional dependence is given \cite{Badger:2013gxa} by
\begin{equation}
  \mathcal{A}^{[D_s]}_{\text{external}\,\in\,S_{[4]}} = \mathcal{A}^{[6]}_{\mathcal{G}} + (D_s-6) \mathcal{A}^{[6]}_{\mathcal{GS}} +(D_s-6)^2 \mathcal{A}^{[6]}_{\mathcal{S}},
  \label{eq:2loopds}
\end{equation}
where $\mathcal{A}^{[6]}_{\mathcal{G}}$ receives contributions from diagrams with only 6d gluons;
$\mathcal{A}^{[6]}_{\mathcal{GS}}$ from diagrams with one closed scalar loop;
$\mathcal{A}^{[6]}_{\mathcal{S}}$ from diagrams with two closed scalar loops. The latter is only possible
for factorized (or ``one-loop-squared'') topologies, i.e. when two loops do not share a propagator (see e.g. fig.~\ref{fig:facttopology}).
Compared to the one-loop case there is no decoupling of 6d gluons and scalars at the level of full amplitudes since the $(D_s-6)$ coefficient
contains diagrams with both 6d gluons and scalars.  Another subtlety is that one is forced to consider two different
``flavors'' of scalars for the evaluation of $\mathcal{A}^{[6]}_{\mathcal{S}}$ since the four-gluon vertex couples
scalars coming from different $D_s$.

In what follows we extend eq.~\eqref{eq:2loopds} to amplitudes with external fermions.
In this case we have to compute partial traces as prescribed by eq.~\eqref{eq:trampl}.
The tensor product structure of the Clifford algebra allows us to accomplish this in two steps:
\begin{equation}
  \trh[D_s\to4]{\,\cdot\,} = \trh[6\to4]{\trh[D_s\to 6]{\,\cdot\,}},
  \label{eq:tracesplit}
\end{equation}
where the traces $\trh[D_s\to4]{\,\cdot\,}$ are defined in the section \ref{subsec:trtw}; and
$\trh[D_s\to6]{\,\cdot\,}$ can be defined similarly with the target $D_s=4$ space replaced
by the $D_s=6$ one. First we trace over all degrees of freedom beyond
$D_s=6$, which can be done once and for all. The remaining trace from $D_s=6$ to $D_s=4$ in general has to be evaluated explicitly (as in eq.~\eqref{eq:explicittrace}).
In order to evaluate
the inner trace we split gamma matrices as
\begin{subequations} \label{eq:tensdecomp2l}
  \begin{align}
    \gm{\mu}{6}{\dt} &= \prescript{(\dtt^\prime)}{}{\mathbb{1}}\otimes \gm{\mu}{6}{8}, \\ \label{eq:tensdecomp2lb}
    \gm{\mu}{D_s-6}{\dt}  &=
    \htw{\mu}{(\dtt^\prime)}\otimes  \prescript{(8)}{}{\Gamma^*},
  \end{align}
\end{subequations}
where $\dtt^\prime = \dt/8$ (cf. \eqns{eq:fourgamma}{eq:highgamma}).
The quark-scalar vertex is thus proportional to $\pres{\Gamma^*}{8}$ since it arises from gluon polarizations beyond six dimensions.

\begin{figure}[t]
    \centering
    \centering
    \subfloat[][]{\includegraphics[height=\Dht]{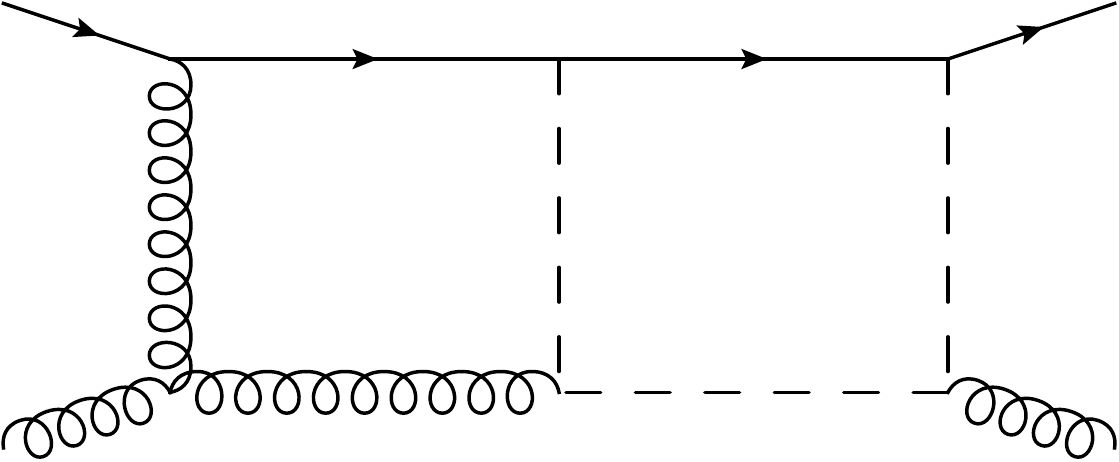}}
    \qquad
    \subfloat[][]{\includegraphics[height=\Dht]{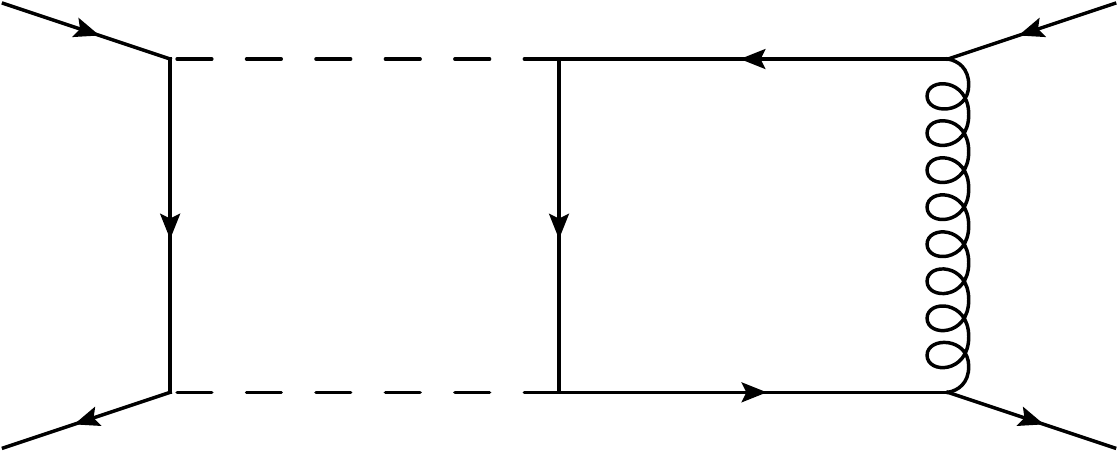}}
    \caption{Diagrams contributing to the $(D_s-6)$ coefficient.}
    \label{fig:Dlin}
\end{figure}

\begin{figure}[h]
    \centering
    \subfloat[][skeleton contraction diagram]{\label{subfloat:sk} \includegraphics[height=12ex]{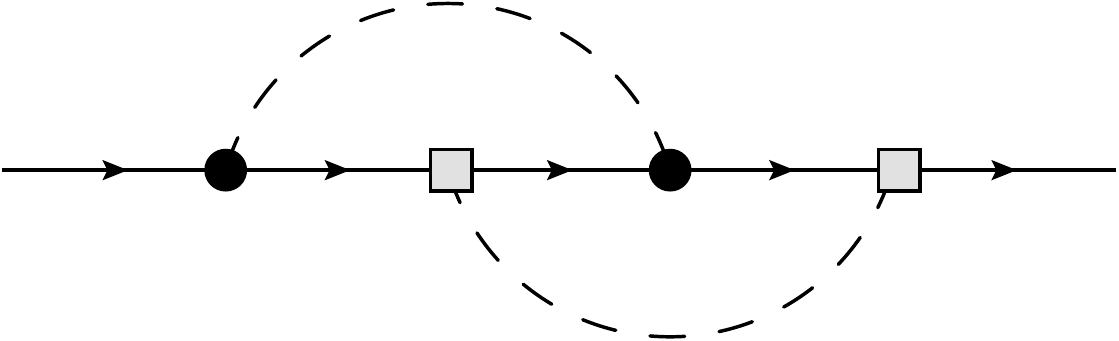}}
    \qquad
    \subfloat[][double box diagram]{\label{subfloat:db} \includegraphics[height=\Dht]{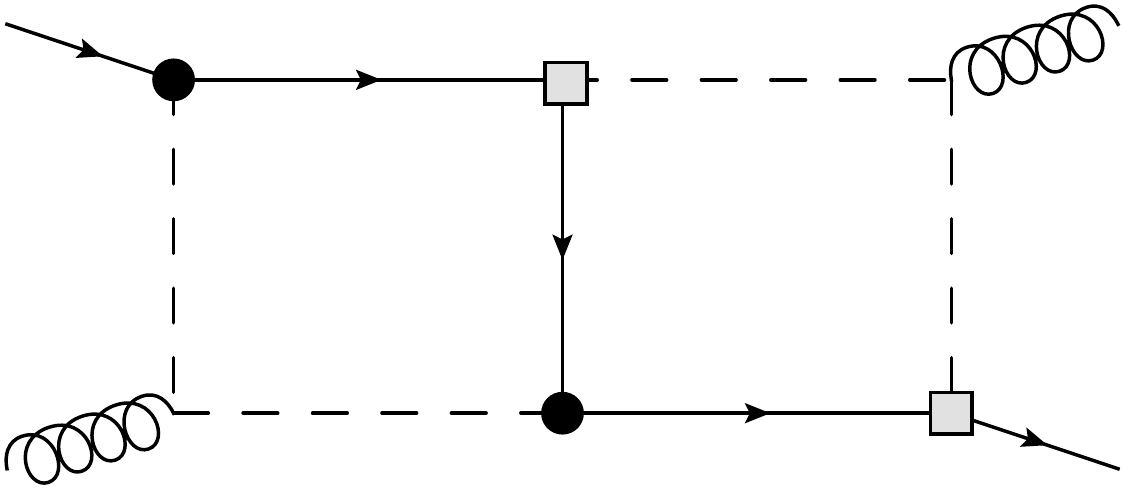}}
    \caption{Diagrams contributing to both $(D_s-6)$ and $(D_s-6)^2$ coefficients. The diagram \protect\subref{subfloat:sk} represents a skeleton diagram,
    where any number of 6d gluons can be inserted on fermion or scalar lines. The diagram \protect\subref{subfloat:db} is an example of such an insertion.
    Quark-scalar vertices originating from equal indices are represented by the same shape.}
    \label{fig:Dmixed}
\end{figure}

The only source of matrices $\Gmds{\mu}{D_s-6}$ that are non-trivial in the $(D_s-6)$ space are interactions of $D_s$-dimensional gluons and quarks. Since traces over odd numbers
of gamma matrices vanish, there are three possibilities:
\begin{enumerate}
  \item \label{iteminternal1} Quarks only couple to 6d gluons, which means that no $\htw{\mu}{}$ insertions are present and the $\trh[D_s\to 6]{\,\cdot\,}$ is trivial.
  \item \label{iteminternal2} Two quark-scalar vertices give $\trh[D_s\to 6]{\,\htw{\mu}{}\htw{\nu}{}\,} = g_{[D_s-6]}^{\mu\nu}$, which contracted with a scalar line contributes a $(D_s-6)$ factor
    (see fig.~\ref{fig:Dlin}).
  \item  Four quark-scalar vertices give
    \begin{multline}
      T^{\mu\nu\delta\sigma} = \trh[D_s\to 6]{\textfrak{h}^{\mu} \textfrak{h}^{\nu} \textfrak{h}^{\delta}\textfrak{h}^{\sigma}} =
      g_{[D_s-6]}^{\mu\sigma} g_{[D_s-6]}^{\nu\delta}-g_{[D_s-6]}^{\mu\delta} g_{[D_s-6]}^{\nu\sigma}+g_{[D_s-6]}^{\mu\nu} g_{[D_s-6]}^{\delta\sigma},
    \end{multline}
    which can be contracted with scalars in two non equivalent ways:
    \begin{subequations}
      \begin{align}
        T_{\mu\nu\delta\sigma}~ g_{[D_s-6]}^{\mu\sigma}g_{[D_s-6]}^{\nu\delta} &= (D_s-6)^2, \\
        \label{eq:Dmixed}
        T_{\mu\nu\delta\sigma}~ g_{[D_s-6]}^{\mu\delta}g_{[D_s-6]}^{\nu\sigma} &= 2(D_s-6)-(D_s-6)^2.
      \end{align}
    \end{subequations}
    The above possibilities are illustrated by fig.~\ref{fig:D2generic} and fig.~\ref{fig:Dmixed} respectively.
\end{enumerate}
In cases \ref{iteminternal1} and \ref{iteminternal2} an additional power of $(D_s-6)$ from a closed scalar loop is possible for factorized topologies.

Overall we see that the $D_s$ dependence matches the general structure of eq.~\eqref{eq:2loopds} and can be expressed as
\begin{equation}
  \mathcal{A}^{[D_s]}_{\text{external}\,\in\,S_{[4]}} = \tilde{\mathcal{A}}^{[6]}_{\mathcal{G}} + (D_s-6) \tilde{\mathcal{A}}^{[6]}_{\mathcal{GS}} +(D_s-6)^2 \tilde{\mathcal{A}}^{[6]}_{\mathcal{S}},
  \label{eq:2loopdsQ}
\end{equation}
where tildes on the RHS remind that the trace $\trh[6\to4]{\cdot}$ is to be evaluated.
Here $\tilde{\mathcal{A}}^{[6]}_{\mathcal{G}}$ is computed by replacing all $D_s$-dimensional gluons with 6d gluons.
All diagrams with both 6d gluons and scalars,
as well as doubled diagrams with alternating scalar lines (as in fig.~\ref{fig:Dmixed}) contribute to $\tilde{\mathcal{A}}^{[6]}_{\mathcal{GS}}$.
The latter follows from eq.~\eqref{eq:Dmixed}.
The $(D_s-6)^2$ coefficient $\tilde{\mathcal{A}}^{[6]}_{\mathcal{S}}$, compared to the pure Yang-Mills case,
receives contributions from generic diagrams with scalars (as in fig.~\ref{fig:D2generic}).
The alternating scalar diagrams contribute with a minus sign which arises from eq.~\eqref{eq:Dmixed}.

Once the $D_s$ dependence is established an FDH or HV amplitude can be obtained by insertion of appropriate values of $D_s$.
The non-trivial difference between the two is given by
\begin{equation}
  \mathcal{A}^{\text{HV}}-\mathcal{A}^{\text{FDH}} = 2\left( 4 \tilde{\mathcal{A}}^{[6]}_{\mathcal{S}} - \tilde{\mathcal{A}}^{[6]}_{\mathcal{GS}}\right) \epsilon +
  4 \tilde{\mathcal{A}}^{[6]}_{\mathcal{S}}\epsilon^2,
\end{equation}
which underlines the known problems with consistency of the ``naive'' FDH scheme beyond one loop \cite{Jack:1994bn,Gnendiger:2017pys,Kilgore:2012tb,Signer:2008va}.

\begin{figure}[h]
    \centering
    \subfloat[][]{\label{subfloat:nf} \includegraphics[height=\Dht]{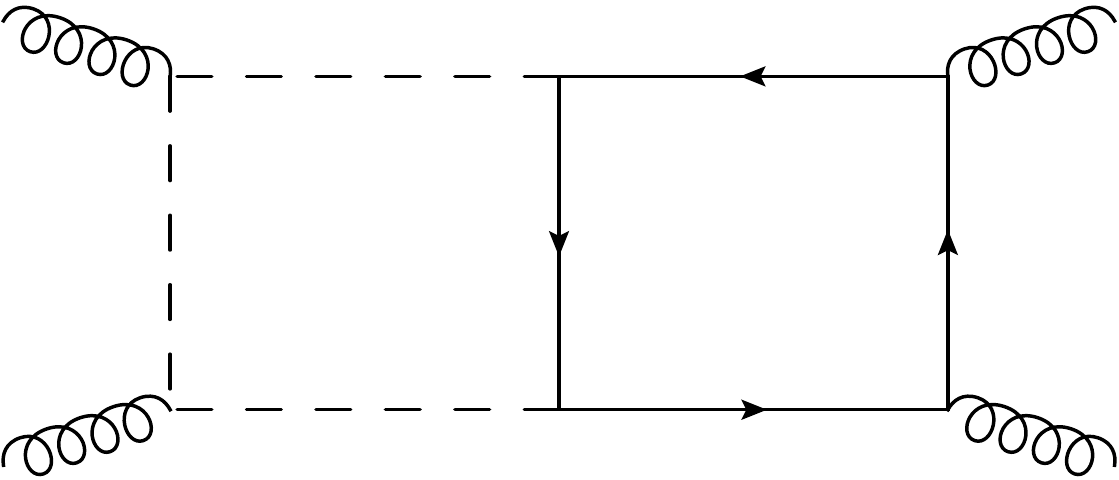}}
    \qquad
    \subfloat[][]{\label{subfloat:nf2} \includegraphics[height=\Dht]{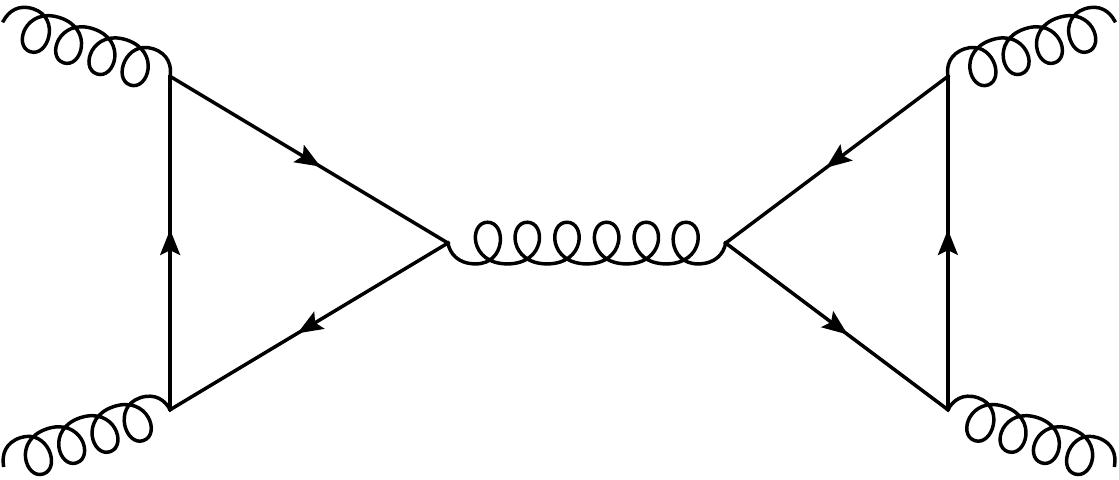}}
    \caption{Examples of diagrams with closed quark loops. The diagram \protect\subref{subfloat:nf} is proportional to the number of flavors and
    contributes to the $(D_s-6)$ coefficient. The diagram \protect\subref{subfloat:nf2} is proportional to the number of flavors squared.}
    \label{fig:2loopnf}
\end{figure}

\begin{subequations}
In a similar way we consider amplitudes with closed quark loops.
When there is one closed quark loop at most two quark-scalar vertices are possible (see e.g. fig.~\ref{subfloat:nf}), which gives at most a linear dependence on $(D_s-6)$:
\label{eq:2loopdsnf}
\begin{equation}
  \mathcal{A}^{[D_s]}_{\text{nf, external}\,\in\,S_{[4]}} = \dtt^\prime \left( \tilde{\mathcal{A}}^{[6]}_{\text{nf}~\mathcal{G}} + (D_s-6) \tilde{\mathcal{A}}^{[6]}_{\text{nf}~\mathcal{S}} \right),
\end{equation}
where a factor $\dtt^\prime$ comes from an additional partial trace over the quark loop.
In case of two closed quark loops (see e.g. fig.~\ref{subfloat:nf2}) no $(Ds-6)$-dimensional index contractions are possible and again we have a factor $\dtt^\prime$
from each quark loop:
\begin{equation}
  \mathcal{A}^{[D_s]}_{\text{nf$^2$, external}\,\in\,S_{[4]}} = \left( \dtt^\prime \right)^2 \mathcal{A}^{[6]}_{\text{nf}^2}.
\end{equation}
\end{subequations}

Summing up, eqs.~\eqref{eq:2loopdsQ} and \eqref{eq:2loopdsnf} together with eq.~\eqref{eq:explicittrace} provide an efficient algorithm for evaluation of dimensionally regularized two-loop amplitudes with
massless or massive quarks employing representations of the minimal dimensionality.

%%%%%%%%%%%%%%%%%%%%%%%%%%%%%%%%%%%%%%%%%%%%%%%%%%%
%%%%%%%%%%%%%%%%%%%%%%%%%%%%%%%%%%%%%%%%%%%%%%%%%%%

\section{Connection to the FDF}
\label{sec:fdf}
The recent four-dimensional (re)formulation (FDF) of
FDH \cite{Fazio:2014xea} provides a computational prescription for
one-loop amplitudes in four dimensions together with
selection rules that have to be evaluated diagrammatically. The FDF scheme is defined with the additional requirement to
remove odd powers of $\mu$ from the integrand. We find that for amplitudes with gluons and
up to one quark pair, the application of
the FDF prescription is equivalent to our decomposition by particle content for one-loop amplitudes in eq.~\eqref{eq:decgsc}. This can be
seen by explicitly comparing selection rule factors and the Feynman rules presented
in Appendix~\ref{sec:impl-deta}. However, for amplitudes involving multiple (massive) quark pairs 
the FDF prescription generates integrands which differ from those of FDH.
In what follows we indicate the source of these spurious integrands, which are not present in our approach.
\begin{figure}[h]%{L}{0.5\textwidth}
  \centering
  \includegraphics[clip,width=23ex]{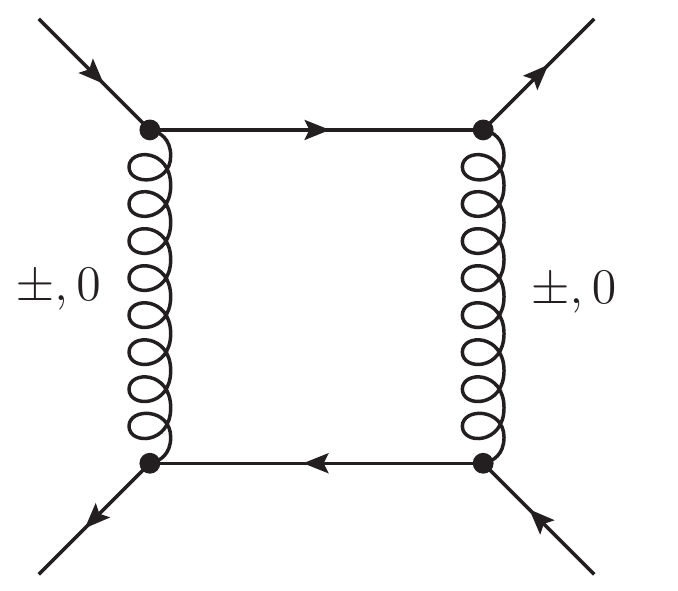}
  \caption{A diagram for the exchange of two ``massive''
    gluons between two quark lines in the FDF. It contains terms that
  are proportional to $\mu^2$, which vanish in our approach. }
  \label{fig:mu2ex}
\end{figure}

As an example, we consider the diagram in Fig.~\ref{fig:mu2ex}, that
is the exchange of two ``massive'' gluons (cf.~Appendix \ref{sec:glue-axial}) between two quark lines. 
Each quark propagator contributes a term $\slashed{l}_{[D-4]} = i \mu \gamma_5$ (c.f. eq.~\eqref{eq:tensdecomp2a}). If both
quark lines are massive%
, there is a non-vanishing contribution of the form
\begin{align}\label{eq:probfdf}
  \langle p_2\vert \gamma^\alpha (i\mu\g5) \gamma^\beta  \vert p_1\rangle ~ \langle
  p_4\vert \gamma^\rho(i\mu\g5) \gamma^\sigma  \vert p_3\rangle ~
  D^{\mathcal{G}}_{\alpha\sigma} D^{\mathcal{G}}_{\beta\rho} ~\sim~ \mu^2 \cdots.
\end{align}
Terms with odd numbers of $\mu$ insertions that combine to even powers
of $\mu$ across multiple lines generically appear in FDF with two
or more massive quark lines. These terms constitute a non-vanishing difference between
integrands in the
FDF and FDH schemes. In our approach these terms vanish due to the
trace over $\dtt$ on each fermion line.

We propose a modification of the FDF prescription to
correctly reproduce one-loop FDH integrands with multiple (massive)
quark lines. Following the notation of ref.~\cite{Fazio:2014xea}, one performs the substitutions
\begin{align}\label{eq:fdfrep}
  g^{\alpha\beta}_{[D_s-4]} &\rightarrow G^{AB}, &
  \gamma^\alpha_{[D_s-4]}&\rightarrow \g5 \Gamma^A,  &
  \ell^\alpha_{[D_s-4]}&\rightarrow i\mu Q^A,
\end{align}
where symbols $\Gamma^A$ do not have a finite-dimensional representation.
The FDF is then
based on the formulation of contraction rules for the objects $G^{AB}$,
$\Gamma^A$ and $Q^A$. Amongst them,
the rule
\begin{equation}
  Q^A\Gamma^A =1
\end{equation}
is derived by applying the substitutions \eqref{eq:fdfrep} to
\begin{align}\label{eq:wrongsr}
  \slashed{\ell}_{[D_s-4]}  \slashed{\ell}_{[D_s-4]} = - \mu^2
  Q^A\Gamma^AQ^B\Gamma^B &\overset{!}{=} -\mu^2.
\end{align}
However the symbols $\Gamma^A$ have not been equipped with a proper algebraic structure and
in general $Q^A\Gamma^AQ^B\Gamma^B=1$ does not imply $Q^A\Gamma^A=1$.
The corrected selection rule drawn from eq.~\eqref{eq:wrongsr} is thus
\begin{align}\label{eq:nsr}
  Q^A\Gamma^AQ^B\Gamma^B &=1.
\end{align}
A further addition to the basic rules of the FDF approach is to give a proper prescription for
contracting indices across different quark lines equivalent to
eq.~\eqref{eq:traceprop}. Then the FDF becomes a valid (re)formulation of
FDH applicable to amplitudes with multiple quark lines.

\section{Conclusions}
\label{sec:concl}

The main result of the paper is a decomposition by particle content
of one- and two-loop QCD helicity amplitudes regularized in HV or FDH schemes
given by eqs.~\eqref{eq:ddependence},~\eqref{eq:nfdecomp}, and eqs.~\eqref{eq:2loopdsQ},~\eqref{eq:2loopdsnf} respectively.
It is valid for any number of massless or massive fermions.
It is formulated using only finite integer-dimensional representations
of tensor and spinor algebras and extends the previous results
obtained for pure Yang-Mills
theories \cite{Bern:1994cg,Cheung:2009dc,Badger:2013gxa}. Compared to
a direct application of dimensional
reconstruction \cite{Giele:2008ve,Ellis:2008ir}, our method requires
the computation in a single $D_s$ dimension, which is determined by the number of non-trivial directions of loop momenta.
This is in particular beneficial for numerical
unitarity
applications \cite{Ossola:2006us,Ellis:2007br,Giele:2008ve,Berger:2008sj,Abreu:2017xsl,Abreu:2017hqn,Badger:2017jhb} and can be exploited in the future for computation of two-loop
amplitudes with external quarks.

To achieve this we formulated a precise notion of external four-dimensional spinor states embedded into higher dimensional spaces in \eqns{eq:main-prescription}{eq:trampl}.
An important consequence is that no spurious integrands such as square
roots of loop momenta contractions are encountered.
This embedding of states of external particles is required for any numerical approach for computation of multi-loop amplitudes
with external quarks.

We found that the FDF prescription \cite{Fazio:2014xea}, applied to one-loop amplitudes with gluons and those with a single
quark pair, is equivalent to a special case of the formula \eqref{eq:decgsc}. In case of multiple (massive) quark pairs however, the FDF scheme contains spurious terms which are not present in our approach.
We provided an improvement of FDF selection rules to make it fully compatible with our approach, therefore with FDH, also in the
case of amplitudes involving multiple quark lines.

Finally, we provided some details of the algorithms implemented in a new version of the \BH{} library, which has been used
for the computation of phenomenologically relevant NLO QCD corrections \cite{Anger:2017glm}.

\acknowledgments{
We are grateful to H.~Ita and F.~Febres Cordero for valuable
discussions. We thank B.~Page for providing helpful remarks on the manuscript.
We thank R.~A.~Fazio, P.~Mastrolia, and W.~J. Torres Bobadilla for discussions
regarding the relation of our approach to the FDF.

The work of F.R.A. is supported by the Alexander von Humboldt
Foundation, in the framework of the Sofja Kovalevskaja Award 2014, endowed by
the German Federal Ministry of Education and Research. 
V.S.'s work is funded by the German Research Foundation (DFG) within the
Research Training Group GRK~2044.}

\appendix

\section{Evaluation of Normalized Partial Traces}
\label{sec:trace-properties}

Here we list some properties of normalized partial traces defined in the section \ref{subsec:trtw}
which we use throughout the paper.

From \eqns{eq:fourgamma}{eq:trdefprop} one deduces that the matrices $\Gmds{\mu}{4}$ only contribute a unit
matrix to the trace over $\dtt$,
\begin{equation}
  \label{eq:trivial-contribution}
  \trh[]{\Gmds{\mu}{4}\cdot\,\ldots\,} = \gamma^{\mu}~\trh[]{\,\ldots\,}.
\end{equation}
The only non-trivial contribution thus comes from partial traces over products of $\Gmds{\mu}{D_s-4}$.
This can be demonstrated by using \eqns{eq:fourgamma}{eq:highgamma}, the fact that  $\left\{ \Gmds{\mu}{4},\Gmds{\nu}{D_s-4} \right\} = 0$, and repeatedly
applying eq.~\eqref{eq:trivial-contribution}.

It follows from eq.~\eqref{eq:trdefprop} that evaluation of partial traces of products of $\Gmds{\mu}{D_s-4}$ is
equivalent to evaluation of normal traces of products of matrices $\gtw{\mu}{}$, which themselves are elements
of a Clifford algebra. Thus all the usual trace properties apply. In particular we have
\begin{subequations}
  \begin{align}
    &\trf[\gtw{\mu_1}{}\cdots\gtw{\mu_n}{}] = 0,\quad n~\text{odd}, \\
    &\trf[\gtw{\mu}{}\gtw{\nu}{}] = g^{\mu\nu}_{[Ds-4]}, \\
    &\trf[\textfrak{g}^{\mu} \textfrak{g}^{\nu} \textfrak{g}^{\delta}\textfrak{g}^{\sigma}] =
          g_{[D_s-4]}^{\mu\sigma} g_{[D_s-4]}^{\nu\delta}-g_{[D_s-4]}^{\mu\delta} g_{[D_s-4]}^{\nu\sigma}+g_{[D_s-4]}^{\mu\nu} g_{[D_s-4]}^{\delta\sigma}.
  \end{align}
\end{subequations}

We note that our rules for computation of normalized partial traces are consistent with the implementation found in \cite{Cullen:2010jv},
where similar objects appear and no connection to finite-dimensional representations is considered.

\section{\texorpdfstring{$\g5$}{g5} in \texorpdfstring{$D_s$}{Ds} Dimensions}
\label{subsec:g5}
The t'Hooft-Veltman/Breitenlohner-Maison \cite{tHooft:1972tcz,Breitenlohner:1977hr} prescription for $\g5$ can
be realized in integer $D_s$ by
  \begin{equation}
    \prescript{(\dt)}{}{\Gamma_{5}} \equiv \prescript{(\dtt)}{}{\mathbb{1}} \otimes  \g5,
  \end{equation}
such that the following relations hold
  \begin{align}
    \left\{\prescript{(\dt)}{}{\Gamma^{\phantom{\mu}}_{5}},\gm{\mu}{4}{\dt}\right\}
    &= 0,  &   \left[\prescript{(\dt)}{}{\Gamma^{\phantom{\mu}}_{5}},\gm{\mu}{D_s-4}{\dt}\right]&=0.
  \end{align}
As a consequence we have 
\begin{align}\label{eq:traceg5}
  \tr{\Gamma_{5} \Gamma^\mu \Gamma^\nu \Gamma^\rho \Gamma^\sigma} \neq 0.
\end{align}
A proper
treatment of the spurious anomalies associated to the fact that $\Gamma_5$
does not anticommute with other Dirac matrices has to be ensured.

\section{Gluon Propagator in Axial Gauge}
\label{sec:glue-axial}
The color-stripped gluon propagator in $D_s$-dimensional ghost-free
axial gauge is given by \cite{Leibbrandt:1987qv}
\begin{align}
  D^{\alpha\beta}_{[D_s]} =
  \frac{i}{\ell_{[D]}^2}\left[-g^{\alpha\beta}_{[D_s]} +
  \frac{\ell^\alpha_{[D]}\eta^\beta_{[D]}+\ell^\beta_{[D]}\eta^\alpha_{[D]}
}{\ell_{[D]}\cdot\eta_{[D]}} - \frac{(\eta_{[D]}^2+\lambda \ell_{[D]}^2)\ell^\alpha_{[D]}\ell^\beta_{[D]}}{(\ell_{[D]}\cdot\eta_{[D]})^2}\right],
\end{align}
with the parameter $\lambda$ and the $D$-dimensional reference
vector $\eta_{[D]}^\alpha$. We choose the reference vector $\eta_{[D]}^\alpha =
i\mu n^\alpha_4$ with $\eta_{[D]}^2=-\mu^2$ , proportional to the unit vector in the fifth dimension
$n_4^\alpha$. We set the parameter $\lambda \rightarrow 0$ and split the metric and loop momentum into
four- and higher-dimensional parts. The above expression then simplifies to
\begin{align}
  D^{\alpha\beta}_{[D_s]} =  \frac{i}{\ell_{[4]}^2-\mu^2}\left[-g^{\alpha\beta}_{[4]}+
    \frac{\ell^\alpha_{[4]}\ell^\beta_{[4]}}{\mu^2}-g^{\alpha\beta}_{[D_s-5]}\right]  \equiv D^{\alpha\beta}_{\mathcal{G},[4]} + D^{\alpha\beta}_{\mathcal{S},[D_s-5]},
\end{align}
where we used the fact that the $(Ds-4)$-dimensional part of the metric can be written as
\begin{align}
  g^{\alpha\beta}_{[D_s-4]} = \sum_{k=4}^{D_s-1}n^\alpha_kn_k^\beta.
\end{align}
The $D_s$-dimensional gluon propagator in axial gauge thus decomposes
into the propagator of a ``massive'' vector particle in four dimensions
and scalar propagators of higher-dimensional components. For on-shell
loop momenta with $\ell_{[D]}^2=0$, the reference vector
$\eta_{[D]}^\alpha = \ell^\alpha_{[4]} - \ell^\alpha_{[D-4]}$ leads to the
same split-up into a ``massive'' four-dimensional gluon and scalars.

The only non-vanishing
contractions of $n_{k\geq 5}$ are with itself via metric
tensors, since external polarizations and momenta as well as the ``massive'' gluon propagator are four-dimensional and the loop momentum
is five-dimensional. As a consequence, scalar lines
can start and end only on quark lines, or otherwise they form a closed loop.
The three-gluon vertex in $D_s$ dimensions,
with two gluons polarized along $n_k$, corresponds to a
scalar-scalar-gluon vertex \cite{Giele:2008ve}
\begin{align}
  V_{\alpha\beta\rho}^{ggg}(p_1,p_2,p_3)
  \epsilon_{1}^{\alpha}n_{k}^{\beta}n_{k}^{\rho}\sim \epsilon_{1}^{\alpha} (p_2-p_3)_\alpha,
\end{align}
where $p_i$ are momenta and $\epsilon_1^\alpha$ is the trivially embedded four-dimensional
polarization.
And a similar argument applies to the scalar-scalar-gluon-gluon
vertex. Thus ``massive'' four-dimensional gluons and
each scalar degree of freedom decouple.

%%%%%%%%%%%%%%%%%%%%%%%%%%%%%%%%%%%%%%%%%%%%%%%%%%%
%%%%%%%%%%%%%%%%%%%%%%%%%%%%%%%%%%%%%%%%%%%%%%%%%%%
\section{Implementation Details}
\label{sec:impl-deta}

In this appendix, we provide some details required for a numerical implementation of the
prescription we found in section \ref{sec:dsdep-1loop} for one-loop amplitudes, see
eq.~\eqref{eq:decgsc}. It allows to compute one-loop QCD amplitudes involving multiple (massive)
quark lines regularized
in the FDH scheme in a compact way. An FDH amplitude is given by the difference
of an amplitude with all $D_s$ dimensional gluons replaced by a four-dimensional
``massive'' gluon in the loop and one
with a scalar particle in the loop, see eq.~\eqref{eq:decgsc}. Furthermore, these amplitudes are defined in
terms of four-dimensional objects with the additional requirement of
only even numbers of $(D_s-4)$-dimensional loop momenta insertions
from quark propagators on each fermion line enforced.

We
propose to keep track of the number of $(D_s-4)$ insertions from quark propagators to
enforce this requirement. One way to achieve this numerically is to use a double copy
of fermionic states and vertices and to adapt the $\mu$-dependent piece of the propagator to switch between upper and lower components, see Feynman rules in Table \ref{tab:fr}. For external and internal fermionic particles, we construct states as described in the section \ref{sec:fermionic-states} with $D_s = 6$. For external particles, only the states with \dttindex{} $j = 0$ are considered,
which projects out terms with even numbers of insertions of $(D_s-4)$-dimensional loop
momenta. One can equally think of this as keeping track of even powers of $\mu$. Note, that this
is not the same as computing an amplitude in $D_s=6$ dimensions.

\ifx
\begin{figure}[h]
  \centering
  \begin{align}
    \vcenter{\hbox{\includegraphics[clip,width=0.2\textwidth]{figures/0q}}}
    &\xrightarrow{D_s\rightarrow 4}   \vcenter{\hbox{\includegraphics[clip,width=0.221\textwidth]{figures/0q_glue}}} -   \vcenter{\hbox{\includegraphics[clip,width=0.2\textwidth]{figures/0q_sc}}}\\
    \vcenter{\hbox{\includegraphics[clip,width=0.2\textwidth]{figures/0q2}}}
    &\xrightarrow{D_s\rightarrow 4}   \vcenter{\hbox{\includegraphics[clip,width=0.2\textwidth]{figures/0q_q}}}\\
    \vcenter{\hbox{\includegraphics[clip,width=0.2\textwidth]{figures/1q}}}
    &\xrightarrow{D_s\rightarrow 4}
    \vcenter{\hbox{\includegraphics[clip,width=0.221\textwidth]{figures/1q_glue}}}
    -
    \vcenter{\hbox{\includegraphics[clip,width=0.2\textwidth]{figures/1q_sc}}}\\
    \vcenter{\hbox{\includegraphics[clip,width=0.2\textwidth]{figures/2q}}}
    & \xrightarrow{D_s\rightarrow 4}     \vcenter{\hbox{\includegraphics[clip,width=0.221\textwidth]{figures/2q_glue}}} -   \vcenter{\hbox{\includegraphics[clip,width=0.2\textwidth]{figures/2q_sc}}}
  \end{align}
  \caption{ The schematic computation of FDH amplitudes in terms of ``massive'' gluon
    and scalar amplitudes with even numbers of $(D_s-4)$-dimensional loop
    momenta insertions. We show the examples of gluon and closed fermion
    diagrams, as well as those with (multiple) quark lines entering the loop.
    The dimension $\dt$ of fermion states in the loop is written between parentheses.
  }
  \label{fig:examplesimpl}
\end{figure}

\fi

An additional simplification can be
exploited in case when quarks are massless. $\g5$ matrices acting on Weyl spinors
preserve representations as opposed to normal slashed matrices.
Hence, spinor products with an odd number of $\g5$ matrices, inserted in place of slashed matrices,
produce a mismatch between representations and have to vanish.
The requirement to have only even numbers of $(D_s-4)$-dimensional loop momentum insertions is thus automatically
fulfilled. In this situation, it suffices to compute with a four-dimensional representation of gamma matrices and the quark propagator 
\begin{align}\label{eq:4dquark}
\prescript{(4)}{}{ S_{\text{q}}(\ell_{[D]})}= \frac{i\left(\prescript{(4)}{}{\slashed{\ell}_{[4]}} + i\mu \g5 + m\right)}{\ell_{[4]}^2-\mu^2-m^2},
\end{align}
with $m=0$. 

A similar simplification applies in case of closed fermion
loops. Odd numbers of \mbox{$(D_s-4)$-}dimensional loop momenta insertions
lead to vanishing traces over odd numbers of $\g5$. Consequently,
the requirement of having even numbers of insertions is always
fulfilled and it
suffices to compute in four-dimensions with the quark propagator of
eq.~\eqref{eq:4dquark}, even with $m\neq0$. 

In the two cases discussed above the simplified quark propagator given
be eq.~\eqref{eq:4dquark} is identical to the one of the FDF approach \cite{Fazio:2014xea}.

\begin{table}[]
\centering
\caption{Color-ordered QCD Feynman rules for the direct computation of
  FDH one-loop amplitudes. Shown are the rules for ``massive'' gluons
  and scalars in the loop. The requirement to have even
  numbers of insertions of higher-dimensional loop momentum along each
fermion line is enforced by construction.}
\label{tab:fr}
\begin{tabular}{l}
\begin{tabularx}{\textwidth}{ll}
    \noalign{\vskip 2.5mm}
    \hline\hline
    \noalign{\vskip 2.5mm}
    \textbf{Propagators}\\
    \noalign{\vskip 2mm}
    \hline
    \noalign{\vskip 4mm}
\includegraphics[clip,width=0.2\textwidth]{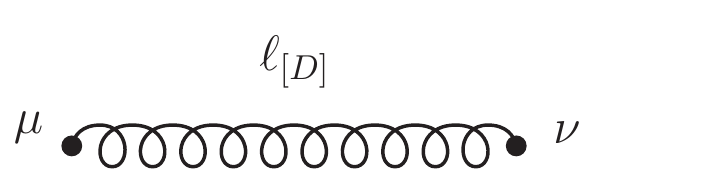}
&$\displaystyle =\frac{i}{\ell_{[4]}^2-\mu^2}\left[-g^{\mu\nu}_{[4]}+
  \frac{\ell^\mu_{[4]}\ell^\nu_{[4]}}{\mu^2}\right]$\\
\includegraphics[clip,width=0.15\textwidth]{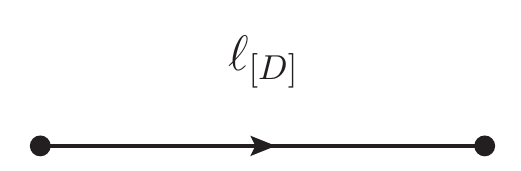} &
$\displaystyle  = \frac{i}{\ell_{[4]}^2-\mu^2-m^2} \left[\prescript{(2)}{}{\mathbb{1}}\otimes
  \left(\prescript{(4)}{}{\slashed{\ell}_{[4]}} +
     m\,\prescript{(4)}{}{\mathbb{1}}\right)  + \pmqty{0 & 1 \\ 1 & 0}
   \otimes i\mu\g5\right] $\\
\includegraphics[clip,width=0.15\textwidth]{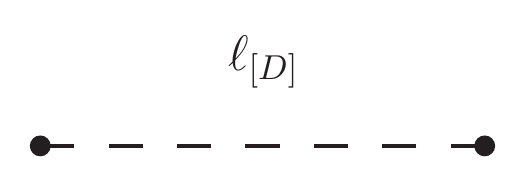} &
$\displaystyle  = \frac{-i}{\ell^2_{[4]}-\mu^2} $\\
    \noalign{\vskip 2.5mm}
    \hline
    \noalign{\vskip 2.5mm}
    \textbf{Vertices}\\
    \noalign{\vskip 2mm}
    \hline
    \noalign{\vskip 4mm}
\raisebox{-.4\height}{\includegraphics[clip,width=0.2\textwidth]{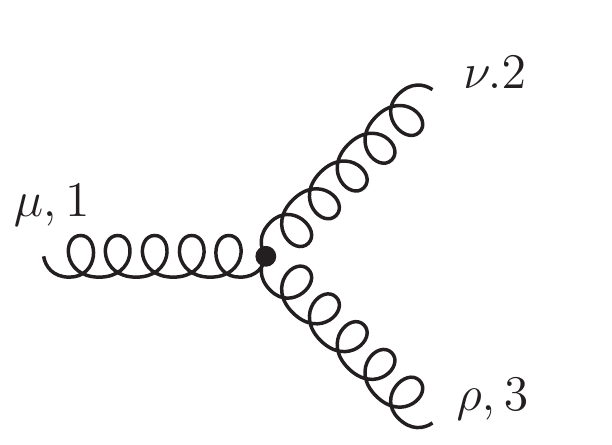}} &
$\displaystyle  = \frac{i}{\sqrt{2}}\left[g_{[4]}^{\mu\nu}(p_{1\,[4]}-p_{2\,[4]})^\rho+g_{[4]}^{\nu\rho}(p_{2\,[4]}-p_{3\,[4]})^\mu+g_{[4]}^{\rho\mu}(p_{3\,[4]}-p_{1\,[4]})^\nu\right]
$\\
\raisebox{-.4\height}{\includegraphics[clip,width=0.2\textwidth]{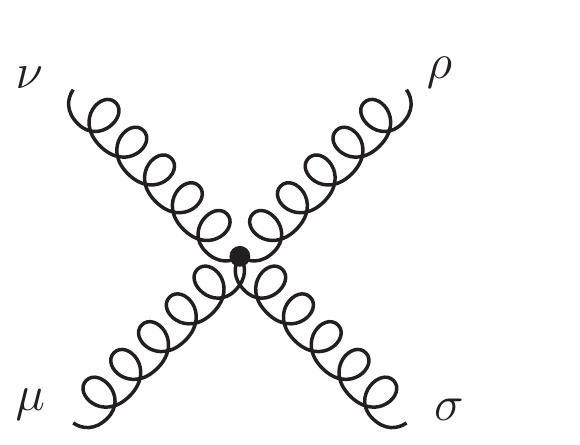}} &
$\displaystyle  = ig_{[4]}^{\mu\rho}g_{[4]}^{\nu\sigma} -
\frac{i}{2}\left[g_{[4]}^{\mu\nu}g_{[4]}^{\rho\sigma}+g_{[4]}^{\mu\sigma}g_{[4]}^{\nu\rho}\right]
$
\end{tabularx}\\
\begin{tabularx}{\textwidth}{llll}
\raisebox{-.4\height}{\includegraphics[clip,width=0.2\textwidth]{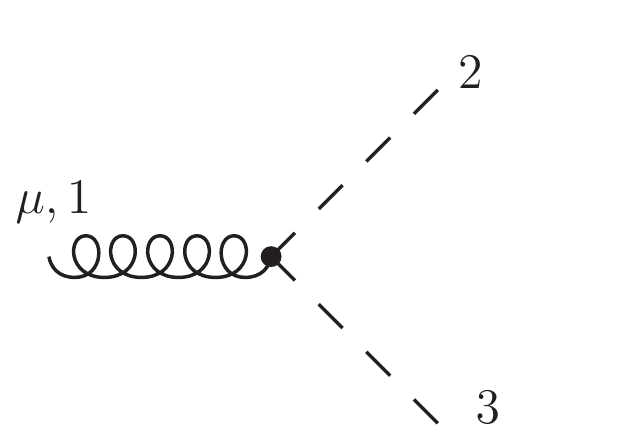}} &
$\displaystyle  =  \frac{i}{\sqrt{2}}  (p_{2\,[4]}-p_{3\,[4]})^\mu$&
\raisebox{-.4\height}{\includegraphics[clip,width=0.15\textwidth]{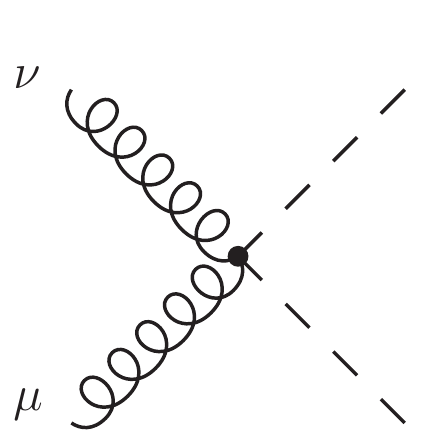}} &
$\displaystyle  =  - \frac{i}{2}   g_{[4]}^{\mu\nu}$\\
\raisebox{-.4\height}{\includegraphics[clip,width=0.15\textwidth]{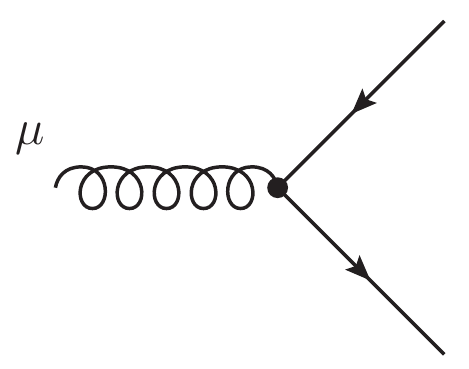}} &
$\displaystyle  = \frac{i}{\sqrt{2}} \prescript{(2)}{}{\mathbb{1}}
\otimes \gamma^\mu $&
\raisebox{-.4\height}{\includegraphics[clip,width=0.15\textwidth]{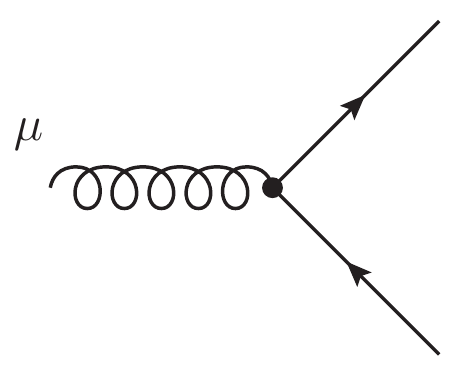}} &
$\displaystyle  = -\frac{i}{\sqrt{2}} \prescript{(2)}{}{\mathbb{1}}
\otimes \gamma^\mu $\\
\raisebox{-.4\height}{\includegraphics[clip,width=0.15\textwidth]{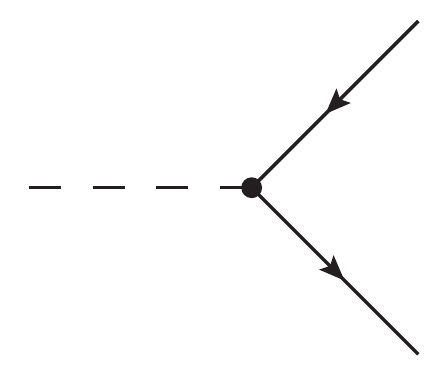}} &
$\displaystyle  = \frac{i}{\sqrt{2}} \prescript{(2)}{}{\mathbb{1}}\otimes\g5$&
\raisebox{-.4\height}{\includegraphics[clip,width=0.15\textwidth]{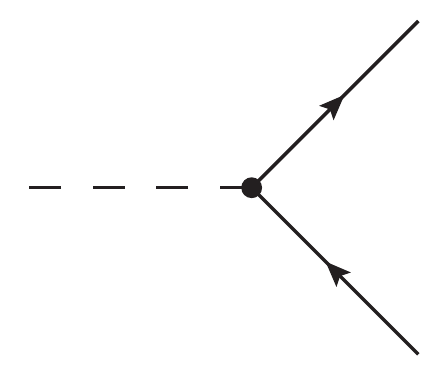}} &
$\displaystyle  = -\frac{i}{\sqrt{2}}\prescript{(2)}{}{\mathbb{1}}
\otimes\g5$\\
    %\noalign{\vskip 4mm}
    %\hline
    %\noalign{\vskip 2.5mm}
    %\textbf{Outgoing Fields}\\
    %\noalign{\vskip 2mm}
    %\hline
    %\noalign{\vskip 3mm}
%\raisebox{-.1\height}{\includegraphics[clip,width=0.15\textwidth]{figures/wfq}} &
%$\displaystyle  = \pmqty{\prescript{(4)}{}{\bar{u}(p)},&0}$&
%\raisebox{-.1\height}{\includegraphics[clip,width=0.15\textwidth]{figures/wfqb}} &
%$\displaystyle  = \pmqty{\prescript{(4)}{}{v(p)} \\ 0}$\\
    \noalign{\vskip 4mm}    
\hline\hline
\end{tabularx}
\end{tabular}
\end{table}

\clearpage

\bibliography{ref}

\end{document}